\documentclass{emulateapj}

\shorttitle{Jets in XRBs} \shortauthors{Pe'er \& Casella}

\newcommand{\beq}{\begin{equation}}
\newcommand{\eeq}{\end{equation}}
\newcommand{\ba}{\begin{array}}
\newcommand{\ea}{\end{array}}

\newcommand{\ee}{\epsilon_{e,0}}
\newcommand{\eB}{\epsilon_{B,-3}}

\def \etal{{\it et al.~}}

\begin{document}
\title{A Model for Emission from Jets in X-ray Binaries: Consequences of a
  Single Acceleration Episode}  

\author{Asaf Pe'er\altaffilmark{1}\altaffilmark{2}}
\author{Piergiorgio Casella \altaffilmark{3}}  

\altaffiltext{1}{Space Telescope Science Institute, 3700 San Martin
  Dr., Baltimore, Md, 21218; apeer@stsci.edu}
\altaffiltext{2}{Giacconi Fellow}
\altaffiltext{3}{Astronomical Institute ``Anton Pannekoek'',
  University of Amsterdam, Kruislaan
  403, 1098SJ, Amsterdam, the Netherlands}

\begin{abstract}

  There are strong evidence for powerful jets in the low/hard state of
  black-hole X-ray binaries (BHXRBs). Here, we present a
  model in which electrons are accelerated once at the base of the
  jet, and are cooled by synchrotron emission and possible adiabatic
  energy losses.  The accelerated electrons assume a Maxwellian
  distribution at low energies and possible energetic power law
  tail. These assumptions yield to a wealth of spectra, which
  we study in details. We identify critical values of the magnetic
  field, and five transition frequencies in the spectra. In
  particular, we show that: (I) the decay of the magnetic field along
  the jet enables, for wide jets, production of flat radio spectra
  without the need for electrons re-acceleration along the jet.  (II)
  An increase of the magnetic field above a critical value of $\sim
  10^5$~G leads to a sharp decrease in the flux at the radio band,
  while the flux at higher frequencies saturates to a constant value.
  (III) For strong magnetic field, the flux decays in the
  optical/UV band as $F_\nu \propto \nu^{-1/2}$, irrespective of the
  electrons initial distribution. (IV) For $B_0 \approx 10^4$~G, the
  X-ray flux gradually steepens. (V) With adiabatic
  energy losses, flat spectrum can be obtained only at a limited
  frequency range, and under certain conditions (VI) For narrow jets,
  $r(x) \propto x^{\alpha}$ with $\alpha < 1/2$, flat radio spectrum
  cannot be obtained.  We provide full description of the spectrum in
  the different scenarios, and show that our model is consistent with
  the key observed properties of BHXRBs.

\end{abstract}


\section{Introduction}
\label{sec:intro}

Observations of the low-hard state of many black hole X-ray binaries
(BHXRBs) reveal a rich broadband spectrum, extending from the radio to
the hard X-rays \citep[for reviews, see, e.g.,][]{Liang98, McR06,
  Vrtilek09}.  Typically, the radio spectrum in the GHz - THz range is
flat ($F_\nu \propto \nu^{-\alpha}$ with $\alpha\sim 0$) or slightly inverted
($\alpha < 0$) \citep{Hynes00, Fender01, Fender06}. This spectrum
extends in some cases up to the infrared range \citep{Chaty03,
  Kalemci05, Migliari07}. At higher energies, from the optical to the
soft X-ray frequencies, the emission is dominated by a thermal
irradiated disk and the companion star, while at even higher energies,
hard X-ray to soft $\gamma$-rays, a power law spectral distribution is
typically observed, with power law index $0.5 \lesssim \alpha \lesssim
1.1$ and variable exponential cutoff \citep[][and references therein]{McR06}.

Two classes of models are often used in interpreting the non-thermal
part of the X-ray spectrum: accretion-based models
\citep[e.g.,][]{Titarchuk94, Esin97, Poutanen98, Coppi99, Zd00,
  Esin01} and alternatively jet-based models \citep{MFF01, MNCFF03,
  MNW05, BRP06, PBR06, Kaiser06}. Indeed, as jets are widely believed
to be the source of the observed non-thermal radio spectrum, it is
possible that they also play a significant role as a source of the
hard X-ray emission \citep[see, e.g.,][]{MiRo99, Fender06}. This idea
had gained support by the correlation found between the observed radio
and X-ray emission \citep{HHCS98,Corbel00,Corbel03, GFP03}.

A model that explains the flat radio spectra observed by the VLBI in
several compact objects as due to emission from jets was suggested by
\citet{BK79}. In this seminal work, \citet{BK79} showed that a flat
radio spectrum can naturally be obtained due to the change of the
plasma conditions along the jet (in particular, the decay of the
magnetic field and the decrease in the particle number density). This
follows from the self absorption of synchrotron photons, which
produces a pronounced spectral break at $\nu_{thick}$, the frequency
below which the optical depth becomes larger than unity. The flat
radio spectrum follows the dependence of this frequency on the
changing plasma conditions along the jet.

In spite of their successes in reproducing the observed flat radio
spectra, jet models are still incomplete. In some of the models
\citep[e.g.,][]{BK79,MFF01,MNW05, BRP06} adiabatic, as well as
radiative energy losses, are assumed to be fully replenished by an
unknown, continuous re-acceleration process along the entire jet.
While re-acceleration can result from internal shocks within the jet
\citep[e.g.,][]{KSS00, SGLC01, JFK08}, there is no a-priory reason to
assume that it replenishes completely both the adiabatic and radiative
energy losses, since the physical origin of these phenomena are
different.  A model developed by \citet{HJ88} considered only
adiabatic, but not radiative energy losses of the electrons.  Energy
losses of relativistic electrons was considered by \citet{GM98}.
However, this model considered only the optically thin part of the
spectrum, and electrons propagation close to the line of sight.  A
comprehensive study of jet emission which includes energy losses was
carried out by \citet{Reynolds82}, although in this work the effect of
self absorption on the electrons energy spectrum was not considered.

A different approach was used by \citet{Kaiser06}, who suggested a
model in which only a single acceleration episode takes place, and
radiation and adiabatic energy losses are not replenished. In this
work, \citet{Kaiser06} pointed out that due to the decay of the
magnetic field along the jet and because of self absorption effects,
flat radio spectra can naturally be obtained.  \citet{Kaiser06}
considered both conical jet in the ballistic case (i.e., neglecting
adiabatic energy losses), and a scenario in which adiabatic energy
losses are considered. In both cases flat radio spectrum could be
obtained, albeit with specific jet geometry (non-conical jet is
required when adiabatic energy losses are considered). However,
fitting the spectrum of Cyg X-1 in this model could only be done with
a very narrow jet, both in the ballistic and adiabatic scenarios.

In the model by \citet{Kaiser06} presented above (as well as some of
the former models), a power law distribution of the accelerated
electrons in the entire energy range was assumed. The break predicted
in the spectrum is thus only due to the change from optically thick
plasma at low energies, to optically thin above $\nu_{thick}$. While
the details of particles acceleration inside jets are uncertain,
models of particles acceleration in shock waves predict that the power
law distribution of electrons exist only at high energies, while a
significant fraction, perhaps most of the electrons maintain low
energy Maxwellian (thermal) distribution \citep{BE87, Axford94, AB06,
  Spit08}. Thus, an inherent peak in the electrons energy distribution
is expected, at typical Lorentz factor $\gamma_{\min}$.  As a result,
a peak in the synchrotron spectrum is expected at frequency
$\nu_{peak}$, due to emission from electrons at the peak of the
Maxwellian distribution. If $\nu_{peak} < \nu_{thick}$ then this peak
is obscured; however, if $\nu_{peak} > \nu_{thick}$, which, as we show
below, is the case for plasma parameters that are plausible for jets
in black hole binaries, then two breaks in the spectrum are expected,
at $\nu_{thick}$ and at $\nu_{peak}$.

In addition to the characteristic low energy cutoff in the electrons
distribution (attributed to electrons at the peak of the Maxwellian)
any physical acceleration mechanism accelerates electrons only up to a
maximal Lorentz factor, denoted here by $\gamma_{\max}$. This
introduces a maximal frequency $\nu_{\max}$ up to which synchrotron
photons can be emitted \citep[see, e.g.,][]{MFF01}. 
%As was pointed out
%by \citet{Kaiser06} and will extensively discussed below, in models of
%emission along the jet, cooling of the energetic electrons results in
%a shift of $\gamma_{\max}$ (hence of $\nu_{\max}$) to lower energies
%as the electrons propagate along the jet. Since the observed spectrum
%is integrated over emission from different regions along the jet,
%inclusion of $\gamma_{\max}$ has an observable effect on the spectrum
%at much lower energies as well.

The uncertainty that exists in the nature of the acceleration process
is reflected in an inherent uncertainty in the distribution of the
energetic electrons. A power law energetic tail is often assumed;
however, in addition to the uncertainty in the fraction of particles
that are accelerated to the energetic tail,
% mentioned above, 
there is
an uncertainty in the spectral index of the power-law accelerated
particles. Fitting data in supernovae remnants
\citep[e.g.][]{Lazendic04} and gamma-ray bursts afterglow \citep{G98b,
  WG99, FW01} reveal power law distribution with a nearly universal
index, $p = 2.2 \pm 0.2$, which is consistent with theoretical
expectations \citep{BE87}. The X-ray spectra of many BHXRBs in
the low/hard state shows photon index that average at $<\Gamma> \sim
1.7$ \citep{McR06}. If interpreted in the framework of the synchrotron
emission model, this implies electron spectral index $p \sim 2.4$, and
in several cases even higher. The power law indices
inferred from fitting synchrotron emission to the X-ray spectra of
BHXRBs are thus consistently higher than those obtained in other
objects.

An additional inherent uncertainty in the study of emission from
BHXRB's is related to the production and structure of the jet. The jet
may be conical, as assumed in \citet{BK79}, or confined. The jet
geometry can strongly affect the radio spectra \citep{Kaiser06}.
Another uncertainty exists in determining the microphysical processes
in the jetted plasma, which governs the evolution of the magnetic
field and the properties of particle acceleration to high energies,
either via shock waves or other mechanisms (e.g., magnetic
reconnection). The energy in the magnetic field is sometimes assumed
to be in equipartition with the energy carried by the electrons
\citep{Longair94}. However, several authors suggested the hypothesis
of Poynting-flux dominated jets \citep[e.g.,][]{ULRLC00, GS06,
  KBVK07}. Indeed, in many fits \citep[e.g.,][]{MNW05, Gal07, Migliari07},
the energy in the magnetic field is kept as a free parameter, and the
fitted value is somewhat higher than equipartition value. As we show below,
the value of the magnetic field affects the entire shape of the
spectra in a non trivial way.

In this paper, we combine the idea of a single acceleration episode
first proposed by \citet{Kaiser06}, with the recent results on the
distribution of accelerated particles.  We study here a model in which
particles are accelerated only once at the base of the jet, and cool
via synchrotron emission and possible adiabatic energy losses as they
propagate along the jet. The magnetic field decays along the jet from
its initial value at the jet base, $B_0$, which is taken here as a
free parameter.  We assume that the acceleration process produces a
low energy cutoff in the energetic electrons energy distribution, in
the form of a low energy Maxwellian, accompanied by a possible
energetic tail. We show that the inclusion of this inherent cutoff
leads to a variety of complex spectras, which are very different than
that obtained by \citet{Kaiser06}.

This paper is organized as follows.  We first introduce the basic
assumptions of our model in \S\ref{sec:model}. We solve the equations
that describe the cooling rate of the energetic electrons in
\S\ref{sec:elec_temporal_behavior}. These equations hold the key to
the rest of the analysis.  We discuss in \S\ref{sec:general_results}
the general properties of the resulting spectra, and show that
qualitatively different results are obtained for wide and narrow jets
(which are defined there).  We then study the resulting spectra in the
different scenarios in details.  In \S\ref{sec:results1} we study the
basic model of synchrotron emission (neglecting adiabatic energy
losses) from a Maxwellian distribution of electrons.  We extend our
study in \S\ref{sec:results2} to include power law distribution of
energetic electrons.  In \S\ref{sec:results3} and \S\ref{sec:results4}
we consider the effect of adiabatic energy losses on the obtained
spectra for initial Maxwellian and power law distribution of the
energetic electrons, respectively.  We then consider in
\S\ref{sec:results5} the scenario of narrow jets.  We summarize and
conclude in \S\ref{sec:summary}, emphasising the key properties of our
model in view of the existing broad-band data sets. Details of the
numerical model that is used in producing the spectra in
Figg. \ref{fig:2} -- \ref{fig:11} are given in appendix
\ref{sec:numerics}.

\section{The model: basic physical assumptions} 
\label{sec:model}

Our basic jet geometry is similar to the one treated by
\citet{Kaiser06}. We consider a jet centered around the $x$ axis, with
rotational symmetry around that axis. Denoting by $x_0$ the position
at the base of the jet, the jet radius with respect to the $x$-axis at
$x \geq x_0$ is a function of $x$ only, $r=r(x)$. In the following, we
assume a parametric dependence $r(x) = r_0 (x/x_0)^{a_{jet}}$. Here,
$r_0$ is the jet radius with respect to the $x$-axis at $x_0$, and
$a_{jet}$ defines the jet geometry: $a_{jet} = 1$ corresponds to a
conical jet. Since the exact shape of BHXRBs jets are not precisely
known, we keep $a_{jet}$ as a free parameter. In the calculation
below, we assume that the observer is located at an angle $\theta^{\rm
  ob} \gg 0 $ to the jet axis. This assumption allows us to
simplify the radiative transfer calculations of synchrotron photons
along the jet.

We assume that the bulk motion Lorentz factor of the plasma inside the
jet is constant, and equals $\gamma_j$. This assumption can only be
justified for a ballistic jet ($a_{jet} =1$). However, it was shown by
\citet{BR74} that the Lorentz factor of the bulk motion of a flow in a
jet is proportional to $p_x^{-1/4}$, where $p_x$ is the pressure of
the external medium that confines the jet.  Assuming a shallow
pressure gradient, the Lorentz factor of the flow is not expected to
vary significantly along the jet. A similar argument holds for
magnetically confined jet.

The plasma in the jet originates from the accreting disk, hence its
density depends on the disk accretion rate, ${\dot M}_{\rm
  disk}$. Following \citet{FB95} \citep[see also][]{MFF01}, we assume
a fraction $q_j$ of the accreting matter to be injected into the jet.
These assumptions imply that the (comoving) energy density at the base
of the jet is given by $u_0 = q_j {\dot M}_{\rm disk} c^2/\pi r_0^2
\gamma_j \beta_j c$, where $\beta_j \equiv (1-\gamma_j^{-2})^{1/2}$ is
the bulk motion velocity, and $r_0 \ll x_0$ assumed.

We consider here an acceleration process that occurs entirely at
$x_0$. This assumption is of course too simplistic. Nonetheless, we
use it here in order to demonstrate the variety of spectra that can
be obtained, and the spectral dependence on the uncertain value of the
magnetic field, the jet geometry and the distribution of energetic
particles. Given the existing uncertainty in the nature of the
acceleration process, we use two models for the accelerated electrons
energy distribution, which we consider as the two extreme cases. 

The first is a (relativistic) Maxwellian distribution, $n_{el}(\gamma)
d\gamma = A \beta \gamma^2 \exp(-\gamma/\theta_{el}) d\gamma$, where
$\gamma$ is the Lorentz factor associated with the electrons random
motion (not to be confused with the Lorentz factor $\gamma_j$ of the
bulk motion jet flow), $\beta = (1-\gamma^{-2})^{1/2}$ is their random
velocity and $A$ is a normalizaton constant whos exact value is
determined below. The characteristic, normalized electrons
temperature, $\theta_{el}$ at the acceleration site $x=x_0$ is
determined as follows: we assume that the plasma in the jet contains
both electrons and protons. The acceleration process (e.g., a shock
wave) dissipates the kinetic energy of the flow, and redistributes the
proton energy between the electrons and the protons. Hence, the
electrons get some fraction $\epsilon_e$ of the equipartition value of
the total (electrons + protons) energy. Therefore, the average Lorentz
factor associated with the electrons random motion is $\gamma_{\min} =
\epsilon_e (m_p/m_e)$. Since in a Maxwellian energy distribution the
average Lorentz factor is related to the temperature via
$\gamma_{\min} = 3 \theta_{el}$, we conclude that the characteristic
electrons temperature at the base of the jet is
\beq
\theta_{el}(x=x_0) \equiv \theta_{el,0} = \left({\epsilon_e \over
    3}\right) {m_p \over m_e} = 612 \ee.    
\label{eq:theta_el}
\eeq
We adopt here and below the standard convention $Q = 10^x Q_x$ in CGS
units. 

Since a large number of pairs is not expected to be created, the total
number density of electrons is similar to the protons number density
in the plasma. At the base of the jet, it is equal to $n_{el, {\rm \,
    tot}} \approx u_0/m_p c^2$.  The normalization of the accelerated
electrons distribution is given by $A = n_{el, {\rm \, tot}} /2
\theta_{el,0}^3$, where the factor 2 comes from integration over the
electrons energy distribution: $\int_1^\infty n_{el}(\gamma) d\gamma =
A \theta_{el,0} k_2(\theta_{el,0}^{-1})$, where
$k_2(\theta_{el,0}^{-1}) \simeq 2 \theta_{el,0}^2$ is Bessel
k-function of second order, and the last equality holds for
$\theta_{el,0} \gg 1$.

The second electrons energy distribution that is considered here is a
power law distribution above $\gamma_{\min}$, while maintaining a
Maxwellian distribution of the electrons population at lower
energies. This is motivated by the theoretical works discussed in
\S\ref{sec:intro}. In this scenario, the power law index of the
accelerated electrons energy distribution $p = -d\log n_{el}(\gamma) /
d\log(\gamma)$ is taken as a free parameter.
The electrons are accelerated up to maximal
Lorentz factor $\gamma_{\max}$, whose value is determined by equating
the acceleration time and the synchrotron loss time (see discussion in
\S\ref{sec:frequencies}). An exponential cutoff in the electrons
energy distribution above $\gamma_{\max}$ is assumed. 
 
The two distributions considered here rely on strong theoretical basis
and are consistent with modeling spectra from jets in gamma-ray bursts
\citep[GRBs; see, e.g., ][]{PW04}. We note though that alternative
acceleration models exist.  E.g., in modeling emission from jets in
Blazars, \citet{CG08} obtained good fits to the data with electron
distribution that is consistent with power law at low energies,
$\gamma_{\min} \gtrsim 1$.

The magnetic field evolves along the jet. We consider here a scenario
in which the magnetic field is dominated by the toroidal component, $B
= B_\phi \propto r^{-1}$. Therefore, $B = B_0(r/r_0)^{-1} = B_0
(x/x_0)^{-a_{jet}}$. The value of the magnetic field at the base of
the jet, $B_0$ is a free parameter. One way to quantify it is by
assuming that the magnetic field carries some fraction $\epsilon_B$ of
the dissipated kinetic energy: $u_B = \epsilon_B u_0$, resulting in
$B_0 = (8 \pi \epsilon_B u_0)^{1/2}$. An equipartition value of the
magnetic field is therefore obtained by setting $\epsilon_B = 1$. 

% It
%should be noted though, that unlike $\epsilon_e$ which is limited from
%above by equipartition value with the protons kinetic energy, the
%magnetic field can originate from the inner parts of the accreting
%disk. As a result, the value of $\epsilon_B$ as defined here can in
%principle be larger than unity. However, as we show below, significant
%flux at radio frequencies is obtained for much lower values of
%$\epsilon_B \approx 10^{-5} - 10^{-2}$. For these values of
%$\epsilon_B$, the energy carried by the magnetic field is only a small
%fraction of the energy carried by the particles.

As the electrons propagate along the jet at $x>x_0$, they cool via
synchrotron emission and possible adiabatic energy losses. In this
work, we do not consider other physical processes (such as, e.g.,
Compton scattering, or production of pairs). This is due to two
reasons: first, for plausible parameters describing jets in BHXRBs,
other phenomena are much less significant \citep[see,
e.g.][]{Kaiser06}. Second, the aim of this paper is to show that
synchrotron emission combined with adiabatic energy losses can lead by
themselves to a very large variety of possible spectra. When adiabatic
energy losses are included, we use the precise formula, which is
correct in the ultra-relativistic as well as the non-relativistic
regimes,
\beq
{\partial \log (\gamma \beta) \over \partial t}  = -{1 \over
  3}{\partial \log(\Delta V) \over \partial t}, 
\label{eq:ad1} 
\eeq
which leads to $\gamma \beta \propto (\Delta V)^{-1/3}$. Here, $\Delta
V$ is the volume element occupied by the particles; since the particles are
assumed to propagate at constant velocity in the $x$ direction, $\Delta
V \propto r(x)^2$. We thus find that adiabatic energy losses result in
$\gamma \beta \propto r^{-2/3}$. For energetic electrons, $\gamma \gg
1$, this formula asymptotes to the more familiar form, $\gamma \propto
r^{-2/3}$.   

In calculating the observed synchrotron flux, one needs to solve the
radiative transfer equation along the line of sight. We use the
standard assumption that the radiation field is isotropic in the
comoving frame of the fluid. In this work we limit ourselves to consider
observer location at high angle to the jet propagation axis. This
enables us to carry these calculations analytically. The results of
the analytical approximations are then checked with the more precise
numerical calculations, whose details are described in appendix
\S\ref{sec:numerics}.  

We calculate the emission in the direction perpendicular to the $x$
axis by splitting the jet into small segments of length $dx$, and
carrying the radiative transfer calculations at each segment
independently.  The surface area of a segment is $2\pi r(x) dx$. The
calculation can thus be considered as a first order approximation,
although the errors are not expected to be large as long as the angle
between the jet axis and the observer, $\theta^{ob} \gg 0$. In the
calculation of the observed intensity below, we further omit the
Doppler factors for the approaching and receding jets, $\delta_\mp =
[\gamma_j (1 \mp \beta_j \cos \theta^{ob})]^{-1}$. 
%
%For $\theta^{ob} =
%\pi/2$, $\delta_\mp = \gamma_j^{-1}$, which is not very different than
%unity as long as $\gamma_j$ is not much larger than unity. 
%
For arbitrary angle to the line of sight $\theta^{ob}$, the Doppler
factors are within the range $2 \gamma_j \leq \delta_\mp \leq (2
\gamma_j)^{-1}$ ($\delta_\mp[\theta^{ob} = \pi/2] = \gamma_j^{-1}$),
and are therefore not much different than unity for mildly
relativistic jets, $\gamma_j \gtrsim 1$.

\subsection{Characteristic frequencies of synchrotron emission from
 electrons at the base of the jet} 
\label{sec:frequencies}

\subsubsection{Maxwellian distribution of the accelerated electrons}
\label{sec:2.1.1}

We first consider a relativistic Maxwellian energy distribution of
electrons at the base of the jet, $n_{el}(\gamma) d\gamma = A \beta
\gamma^2 \exp(-\gamma/\theta_{el,0})$, with a normalized temperature
given by Eq. \ref{eq:theta_el}. In calculating the characteristic
break frequencies of the observed emission in our model, we use the
same parameters values that were used by \citet{MFF01} in fitting the
broad band spectrum of XTE J1118+480. This gives possible values of
the free model parameters that can be used for illustrative purposes.

We therefore assume a disk accretion rate ${\dot M}_{\rm disk} = 3 \times
10^{-8} {\, \rm M_\odot \, yr^{-1}}$ around a $M_{BH} = 6 {\, \rm
  M_\odot}$ black hole, an efficiency in matter injection into the jet
$q_j = 10^{-2}$ and bulk motion Lorentz factor of the matter inside the
jet $\gamma_j = 2$. We take the jet base to be at distance $x_0 = 45
r_s$ from the black hole, and the jet radius at its base to be $r_0 = 10
r_s$, where $r_s$ is the Schwarzschild radius. With these
assumptions, the energy density at the base of the jet is
\beq
u_0 = {q_j {\dot M}_{\rm disk} c \over \pi r_0^2 \gamma_j \beta_j} =
{10^{26} \over r_0^2} {\bar u_0} \;  {\rm erg \, cm^{-3}},
\label{eq:u_0}
\eeq  
where ${\bar u_0} = q_{j,-2} ({\dot M}_{\rm disk}/10^{7.5}{\rm M_\odot
  \, yr^{-1}}) (\gamma_{j}/2)$ is a dimensionless quantity, whose
value depends on the flow parameters. For the nominal values
taken here, the normalization of the electrons distribution, $A_0
\equiv A(x=x_0) = 5\times 10^{5} \; {\bar u_0} \, r_{0,1}^{-2}\,
\ee^{-3} {\rm \, cm^{-3}}$, 
%\footnote{the subscript ``0''
%  is inserted here and below to denote that the normalization is taken at
%  $x=x_0$; thus $A_0 = A(x=x_0)$.} 
and the magnetic field at the base
of the jet, $B_0 = 9\times 10^4 \, {\bar u_0}^{1/2}\, r_{0,1}^{-1}\,
\eB^{1/2}$~G are readily determined. Here, $r_{0,1} = r_0/ 10 r_s$.

The peak frequency of the spectrum is determined by synchrotron emission
from electrons at the peak of the Maxwellian. At the base of the jet,
\beq
\ba{lcl}
\nu_{peak,0} & \equiv & \nu_{peak}(x=x_0) 
= {3 \theta_{el,0}^2 \over 4 \pi} {q B_0 \over m_e c} \nonumber \\
& \simeq & 
1.4 \times 10^{17} \; {\bar u_0}^{1/2} \, r_{0,1}^{-1}\, \ee^2 \,
\eB^{1/2} \, {\rm Hz}. 
\ea
\label{eq:nu_p}
\eeq
Synchrotron emissivity from a Maxwellian distribution of electrons was
calculated by \citet{JH79}. For $z \equiv 3\nu/2\nu_{peak} \lesssim
1$, it was found that $j_\nu(z\lesssim 1) = (4/9) (A \theta_{el}^3 q^3
B_0/m_e c^2) (z/2)^{1/3}$, while for $z \gg 1$ the flux decays exponentially,
$j_\nu (z \gg 1) = (\pi/4) (A \theta_{el}^3 q^3 B_0 z/m_e c^2)
\exp[-(3/2)(2z)^{1/3}]$.

For a propagation direction perpendicular to the $x$ axis, the
photons travel distance $r(x)$ before escaping the jet.  The optical
depth to synchrotron self absorption of these photons is given by
\citep[e.g.,][]{Rybicki79}
\beq
\tau_\nu = {j_\nu c^2 r(x) \over 2 \nu^2 m_e c^2 \theta_{el}}.
\label{eq:tau}
\eeq

The break frequency is defined as the frequency below which the
optical depth is larger than unity, $\nu_{thick} =
\left.\nu\right|_{\tau_\nu = 1}$. Taking $\nu_{thick} < \nu_{peak}$
(i.e., $z < 1$) at the jet base, using Eqs. \ref{eq:nu_p} and
\ref{eq:tau}, one finds that at the base of the jet
\beq
\ba{lcl}
\nu_{thick,0} & \equiv & \nu_{thick}(x=x_0) \nonumber \\
& = & \left( {2 \over 9} {A
    \theta_{el,0}^2 q^3 B_0 r_0 \over 
    m_e^2 c^2}\right)^{3/5} \left({3 \over 4 \nu_{peak,0}}\right)^{1/5}
\nonumber \\
& = & 
\left( {8\pi^2 \over 9^3} {u_0^4 q^8 r_0^3 \epsilon_B \over m_e^5
    m_p^3 c^{11}} \right)^{1/5} {1 \over \theta_{el,0}} \nonumber \\
& \simeq & 2.1 \times
10^{13} \; {\bar u_0}^{4/5} \, r_{0,1}^{-1}\, \ee^{-1} \, \eB^{1/5} \,
{\rm Hz}.
\ea   
\label{eq:nu_break}
\eeq
The condition $\nu_{peak,0} \geq \nu_{thick,0}$ can thus be written as a
requirement on the energy carried by the electrons,
\beq
\ee \geq 5.3\times 10^{-2}  \; {\bar u_0}^{1/10} \, \eB^{-1/10}.
\label{eq:nu_p_break}
\eeq
The weak dependence on the parameters of the flow and on the magnetic
field, implies that for a wide range of parameters, the right hand
side of Eq. \ref{eq:nu_p_break} is smaller than unity. Thus, for
near equipartition value of the energy carried by the electrons $\ee
\lesssim 1$, the peak emission frequency at the jet base is above the
break frequency.

\subsubsection{Power law distribution of the accelerated electrons}
\label{sec:2.1.2}

Our second acceleration scenario considers power law distribution of
the energetic electrons above the peak of the Maxwellian:
$n_{el}(\gamma) d\gamma \propto \gamma^{-p}$ for $\gamma_{\min} <
\gamma < \gamma_{\max}$.  The power law index $p$ of the electrons
energy distribution is taken as a free parameter. We estimate the
maximum Lorentz factor of the accelerated electrons by equating the
acceleration time, $t_{acc} \simeq E/c q B_0$, with the synchrotron
cooling time, $9 m_e^3 c^5 / 4 q^4 B_0^2 \gamma$, to obtain
$\gamma_{\max} = (3/2) m_e c^2 / (q^3 B_0)^{1/2}$. Synchrotron
emission from these electrons is expected at
\beq
\nu_{\max,0} = {3 \over 4 \pi} {\gamma_{\max}^2 q B_0 \over m_e c} = {27
  \over 16 \pi}{m_e c^3 \over q^2} = 5.7 \times 10^{22} \; {\rm Hz}.
\label{eq:nu_max}
\eeq
Note that this value is independent on the strength of the magnetic
field at the acceleration site.

The minimum Lorentz factor of the power law distribution,
$\gamma_{\min}$ is determined self consistently once the power law
index $p$, the maximum Lorentz factor $\gamma_{\max}$ and the total
number and energy densities of the accelerated electrons are known
(see further explanation on the numerical model in
\S\ref{sec:numerics}). Below $\gamma_{\min}$, a Maxwellian
distribution is assumed, with temperature $\theta_{el,0} =
\gamma_{\min} /2$, chosen such that the low energy Maxwellian smoothly
connects to the power law distribution at higher energies.  The
inclusion of a low energy Maxwellian distribution implies that the
results derived above in Eqs. \ref{eq:nu_p} --
\ref{eq:nu_p_break} hold in this scenario as well.

\subsection{Observed flux  from a jet segment} 
\label{sec:flux}

For an observer located at high angle to the $x$-axis, a jet segment
has a surface area $2 \pi r(x) dx$. The flux observed from a
jet segment of length $dx$ is thus
\beq
dF_\nu = {r(x)^2 \over 2 d^2} {j_\nu \over \tau_\nu}(1-e^{-\tau_\nu}) dx,
\label{eq:dF_nu}
\eeq
where $d$ is the distance, and we omitted the dependence
on the Doppler factors.  

For a Maxwellian distribution of electrons, the observed spectrum from
a jet segment can therefore be described by a broken power law with
two characteristic frequencies: for $\nu \ll \nu_{thick}$, $\tau_\nu
\gg 1$, and
\beq
dF_\nu(\nu \ll \nu_{thick}) \simeq {r(x)^2 \over 2 d^2} {j_\nu \over
  \tau_\nu} dx = {r(x) \over d^2} m_e \theta_{el} \nu^2 dx,
\label{eq:F_nu_thick}
\eeq
is proportional to $\nu^2$.  For $\nu_{thick} \lesssim \nu \lesssim
\nu_{peak}$, the optical depth $\tau_\nu \lesssim 1$ and
\beq
\ba{lcl}
dF_\nu (\nu_{thick} \lesssim \nu \lesssim \nu_{peak}) & \simeq & {r(x)^2
  \over 2 d^2} j_\nu dx \nonumber \\ 
& = & {2 r(x)^2 \over 9 
  d^2} {A \theta_{el}^3 q^3 B \over m_e c^2} \left({3 \nu \over 4
    \nu_{peak}}\right)^{1/3} dx,
\ea
\label{eq:F_nu_thin}
\eeq
is proportional to $\nu^{1/3}$.  For $\nu \gtrsim \nu_{peak}$, the
emissivity $j_\nu$ decays exponentially, and therefore an
exponential cutoff in the observed flux is expected.

In this work we focus on a scenario in which $\epsilon_e$ is close to
equipartition. However, we note that if the electrons carry a much
smaller fraction of the energy such that the condition in Eq.
\ref{eq:nu_p_break} is not fulfilled, then $\nu_{thick,0} >
\nu_{peak,0}$.  Moreover, even if $\epsilon_e \simeq 1$, as we will
show below both $\nu_{peak}$ and $\nu_{thick}$ can vary along the jet
in different ways. As a result, there exists a transition radius
$x_{trans}$ above which $\nu_{thick}(x>x_{trans}) >
\nu_{peak}(x>x_{trans})$. Once this occurs, the emission from a jet
segment is composed of a thick part below $\nu_{thick}$ where
$dF_\nu \propto \nu^2$, and an exponential cutoff at higher
frequencies.

When power law distribution of electrons is considered, in order to 
calculate the emissivity and flux one needs to specify the
fraction of electrons that are accelerated to the high energy tail
above the Maxwellian. For demonstration purposes, we consider here
complete acceleration to the energetic tail, while maintaining the low
energy Maxwellian and high energy cutoff in the distribution. Thus,
the power law distribution smoothly connects to the peak of the
Maxwellian. In this case, calculation of the frequency dependence of
the flux in Eqs. \ref{eq:dF_nu} -- \ref{eq:F_nu_thin} hold (A
slight modification though exist in the overall normalization of the
flux).  

We can thus conclude that there are two main differences in the
emissivity from a jet segment in a power law scenario and the pure
Maxwellian distribution considered above.  The first difference is
that standard calculations \citep[e.g.,][]{Rybicki79} show that above
$\nu_{peak}$ (and below $\nu_{\max}$), there is a power law decrease
of the flux, with power law index $(p-1)/2$. An exponential cutoff
exists only above $\nu_{\max}$, while in a pure Maxwellian
distribution an exponential cutoff exists already above
$\nu_{peak}$.

The second difference is relevant if $\nu_{thick}(x) > \nu_{peak}(x)$.
For power law distribution of electrons, the thick part of the
spectrum ($\nu < \nu_{thick}$) is characterized by $F_\nu \propto
\nu^{5/2}$ \citep[e.g.,][]{Rybicki79}, while a power law $F_\nu
\propto \nu^{(p-1)/2}$ is kept above $\nu_{thick}$.  Here, however, we
expect a somewhat more complex spectral shape at low frequencies $\nu
< \nu_{thick}$: for $\nu \lesssim \nu_{thick}$, $F_\nu \propto
\nu^{5/2}$, while at much lower frequencies, the flux is $F_\nu
\propto \nu^2$ due to the inclusion of a low energy Maxwellian.

\subsection{Electrons energy loss along the jet}
\label{sec:elec_temporal_behavior}

%At any given instance, the total observed flux is calculated by
%integrating the flux emitted from different segments along the
%jet. 
As the electrons propagate along the jet, they cool by
synchrotron emission, as well as possible adiabatic energy losses. As
a result, the characteristic frequencies derived above for $x=x_0$
vary along the jet. We calculate in this section the change in the
electrons energy distribution due to synchrotron and adiabatic
cooling.

As they propagate along the jet, the electrons cool via synchrotron
emission, and their momentum decreases at a rate $\left.d(\gamma
  \beta)/dt\right|_{\rm sync} = -(4/3) c \sigma_T/(m_e c^2) u_B
(\gamma \beta)^2$.  Here, $\sigma_T$ is Thomson's cross section and
$u_B = B^2/8 \pi$ is the energy density in the magnetic field. Along
the jet, the magnetic field decays as $B(x) = B_0
(x/x_0)^{-a_{jet}}$. Since the bulk velocity of the electrons
propagation is taken to be constant, the electrons position along the
jet ($x$) axis is related to the (comoving) time they spend in the jet
by $x = \gamma_j \beta_j c t$; thus, the electrons reach the
acceleration site (at the base of the jet) at time $t_0 = x_0/\gamma_j
\beta_j c$.

Adiabatic energy losses (see Eq. \ref{eq:ad1}) are quantified by
$\left.d(\gamma \beta)/dt\right|_{\rm ad} = -(2/3) (a_{jet}/t) \gamma
\beta$. Considering both synchrotron and adiabatic energy losses, the
electrons momentum decay (in the bulk motion comoving frame) is
governed by \citep{Longair94, Kaiser06}
\beq
{d (\gamma \beta) \over d t} = - {4 \sigma_T \over 3 m_e c} {B_0^2 \over 8
\pi} \left({ t \over t_0 }\right)^{-2 a_{jet}} (\gamma \beta)^2 - {2 a_{jet}
  \over 3 t} (\gamma \beta).
\label{eq:elec_decay} 
\eeq
The solution to Eq. \ref{eq:elec_decay} was derived by
\citet{Kaiser06},
\beq
\gamma \beta(t) = \frac{\gamma \beta(t_0)\times (t/t_0)^{-2 a_{jet}/3}}{1 + 
{\sigma_T B_0^2 t_0 \over 6 \pi m_e c (8 a_{jet}/3
-1)} \gamma \beta (t_0) \left[ 1 - (t/t_0)^{1-8 a_{jet}/3} \right]}
\label{eq:elec_energy1}
\eeq
If adiabatic energy losses are neglected, the second term on the right
hand side of Eq. \ref{eq:elec_decay} drops, and the solution is
\beq
\gamma \beta(t) = \frac{\gamma \beta(t_0)}{1 + 
    { \sigma_T B_0^2 t_0 \over 6 \pi  m_e c (2 a_{jet} -1) }  
    \gamma \beta (t_0) \left[ 1 - 
    (t/t_0)^{1-2 a_{jet}}\right]}.
\label{eq:elec_energy2}
\eeq

Equation \ref{eq:elec_energy2} reveals an important result, which was
first derived by \citet{Kaiser06}: if adiabatic energy losses are not
included, for $a_{jet} > 1/2$, the Lorentz factor of the electrons
asymptotes at $t \rightarrow \infty$ to a value larger than
unity. This implies that some part of the electrons initial energy is
maintained and not radiated. The origin of this counter-intuitive
result is the decay of the magnetic field along the jet. It provides a
natural, possible mechanism for maintaining the electrons (asymptote)
temperature, as is required by, e.g., the model of \citet{BK79},
without the need for particle re-acceleration along the jet. This fact
also allows us to define wide jets in this case as jets for which
$a_{jet} > 1/2$.

\begin{figure}
\plotone{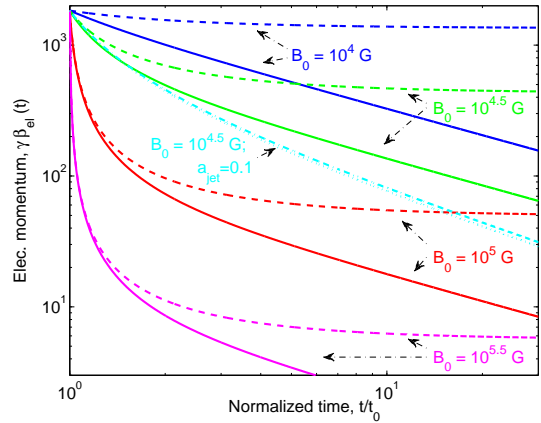}
\caption{ Temporal decay of the electrons momentum, $\gamma\beta(t)$
  as a function of time. Shown are adiabatic + synchrotron (solid
  lines, eq. \ref{eq:elec_energy1}) and pure synchrotron (dashed
  lines, eq. \ref{eq:elec_energy2}) scenarios. Four values of magnetic
  field are considered (from top to button), $B_0 = 10^4, 10^{4.5},
  10^5, 10^{5.5}$~G. The dotted lines (light blue) show the decay of
  the electrons energy for narrow jet, $a_{jet} = 0.1$ and $B_0 =
  10^{4.5}$~G. In all the other cases we consider conical jet
  ($a_{jet} = 1$).  The values of the flow parameters used in
  determining $t_0$ are the same as those used in \S\ref{sec:2.1.1},
  resulting in $t_0 = x_0/\gamma_j \beta_j c = 1.5 \times 10^{-3}$~s,
  and initial electrons momentum $\gamma \beta_{el} (t_0) = m_p/m_e =
  1836$. For pure synchrotron and $a_{jet} > 1/2$, the momentum
  asymptotes at $t \gg t_0$ to a finite value, while for synchrotron
  and adiabatic losses, at $t \gg t_0$, $\gamma \beta (t) \propto
  t^{-2 a_{jet}/3}$. An important result is that the initial rapid
  decay occurs on time scale $t/t_0 \lesssim few$. See text for
  details.  }
\label{fig:temporal}
\end{figure}

We present in Fig. \ref{fig:temporal} several examples of the
electrons momentum temporal decay due to synchrotron and adiabatic
energy losses (Eq. \ref{eq:elec_energy1}) and pure synchrotron
cooling (Eq. \ref{eq:elec_energy2}). In producing the results, we
consider a conical jet ($a_{jet}=1$), and four different values of the
magnetic field, $B_0 = 10^4, 10^{4.5}, 10^5, 10^{5.5}$~G. An
additional plot illustrates the scenario of a narrow jet ($a_{jet} =
0.1$) and $B_0 = 10^{4.5}$~G.  Asymptotic behavior of the electrons
momentum at late times, $t \gg t_0$, is clearly seen (and easily
derived from Eqs. \ref{eq:elec_energy1},
\ref{eq:elec_energy2}). When adiabatic energy losses are considered,
for wide jets, $a_{jet} > 3/8$ the electrons momentum decays
asymptotically as $\gamma \beta \propto t^{-2 a_{jet}/3}$. For pure
synchrotron cooling and $a_{jet} > 1/2$, the electrons momentum
reaches a constant value. Interestingly, we find from Eqs.
\ref{eq:elec_energy1} and \ref{eq:elec_energy2} that in narrow jets,
$a_{jet} < 3/8$, the electrons asymptotic decay law is similar in both
scenarios, i.e., independent on the inclusion of adiabatic energy
losses. In this case, for $t \gg t_0$, both scenarios result in
electrons momentum asymptotic decay law $\gamma \beta \propto t^{2
  a_{jet} -1}$.
 
For strong value of the magnetic field, the asymptotic behavior of the
electrons momentum is reached after a very rapid decay of the
electrons initial energy, or momentum. This rapid decay results from
an extensive synchrotron emission in the strong magnetic field close
to the jet base.  The results presented in Fig. \ref{fig:temporal}
illustrate an important point that will be extensively used in the
analysis in the following sections: in a strong magnetic field, the
initial rapid decay in the electrons momentum takes place on a very
short time scale, $t/t_0 \lesssim few$, which is translated to very
short spatial scale - the rapid cooling (and most of the radiation)
occurs very close to the jet base.

This enables us to obtain an analytical approximation to the observed
flux by splitting the jet into two separated regimes.  We first
calculate the emissivity during the rapid cooling phase, which takes
place close to the jet base, hence the magnetic field can be
approximated as constant during this phase.  At a second step, we
calculate the emissivity during the rest of the electrons propagation
along the jet, assuming that the electrons temperature evolution
follows its asymptotic behavior.  The approximate analytical results
are compared with the exact results obtained numerically.

\section {General Properties of the spectra}
\label{sec:general_results}

As we will show in \S\ref{sec:results1} -- \S\ref{sec:results5},
a wealth of spectra can be obtained, whos details depend on the
various model assumptions.  In particular, we will show that the
observed spectra is very sensitive to the strength of the magnetic
field. This dependence is found to be highly non-trivial.  In
addition, both the jet geometry and the initial particle distribution
significantly affect the obtained spectrum.  Therefore, before
discussing the various possibilities in details, we first discuss in
this section some general properties of the emission along the
jet. Here we focus on global properties of the observed spectra, which
are direct consequences of our models assumptions as stated in
\S\ref{sec:model}.  The detailed spectra expected in the various
scenarios will be discussed in the following sections.

\subsection{Critical values of the magnetic field at the base of the
  jet}
\label{sec:B_critical}
 
We showed in \S\ref{sec:elec_temporal_behavior} that for strong
magnetic field at the jet base, the electrons temporal behavior can be
separated roughly into two distinctive regimes (see figure
\ref{fig:temporal}).  At the first stage, the electrons rapidly cool
due to synchrotron emission. This stage lasts a very short duration,
of the order $t/t_0 \lesssim few$. Following the rapid energy loss, at
a second stage the electrons cooling rate asymptotes to its
terminal value. If only synchrotron emission is considered and
$a_{jet} > 1/2$, then the electrons maintain their energy, and their
Lorentz factor, or momentum, becomes time independent, $\gamma \beta
(t) \propto t^0$ at $t \gg t_0$. If adiabatic energy losses are
considered, then for $a_{jet} > 3/8$ the electrons momentum decays at
late times $t \gg t_0$ in accordance to $\gamma \beta (t) \propto
t^{-2 a_{jet}/3}$. For lower values of $a_{jet}$, both scenarios yield
a similar result, $\gamma \beta (t) \propto t^{2 a_{jet} -1}$.

The fact that the rapid decay phase exists only very close to the jet
base, in regions where the magnetic field is approximately constant
(and equals $B_0$), combined with the fact that the accelerated
electrons distribution has characteristic energies, enables one to
define critical values for the strength of the magnetic field at the
jet base. As we will show below, these critical values represent
qualitative transitions in the resulting spectra. The critical values
are define as follows.

If the acceleration process produces a Maxwellian energy distribution,
then the energetic electrons have a characteristic Lorentz factor,
$\gamma_{\min} (t_0) = 3 \theta_{el,0}$. Due to the exponential decay
nature of this distribution at high energies, there are effectively no
electrons with Lorentz factor $\gamma \gg \gamma_{\min}$. The cooling
rate depends both on the electrons energy, or momentum and on the
strength of the magnetic field at the jet base, $B_0$
(see Eqs. \ref{eq:elec_decay} -- \ref{eq:elec_energy2}). One can
therefore conclude that for very weak magnetic field, the rapid decay
phase is insignificant and the entire electrons distribution is
effectively unchanged during this phase. This case is illustrated in
Fig. \ref{fig:temporal}, for $B_0 = 10^4$~G.  If, on the other hand,
the magnetic field is stronger than a certain value, the initial
electrons cooling is significant.

This fact allows us to define a critical value of the magnetic field
at the jet base, as a value above which electrons at the peak of the
Maxwellian distribution significantly lose their energy by synchrotron
emission during the initial rapid cooling phase. For the pure
synchrotron case (neglecting adiabatic energy losses) and wide jets
($a_{jet} > 1/2$), this critical value is formally obtained by taking
the limit $t \gg t_0$ in Eq. \ref{eq:elec_energy2}, and equating
the second term in the denominator with unity,
\beq 
\ba{lcl} 
B_{cr,1} & = & 1.6 \times 10^4 \, (2 a_{jet} -1)^{1/2} \nonumber \\
 & & \times \ee^{-1/2} \,
(\gamma_j/2)^{1/2} \, (x_0/45 r_s)^{-1/2} \; {\rm G}, \nonumber \\
\epsilon_{B,cr,1} & = & 3\times 10^{-5} \, (2 a_{jet} -1) \nonumber \\
& & \times \ee^{-1}
\, (\gamma_j/2) \, {\bar u_0}^{-1} \, r_{0,1}^{2} \, (x_0/45
r_s)^{-1}.  
\ea
\label{eq:B_cr_1}
\eeq
While the results in Eq. \ref{eq:B_cr_1} are formally derived for
the pure synchrotron scenario in wide jets, one finds that the time
scale of the initial, rapid cooling is similar when adiabatic energy
losses are included (see Eqs. \ref{eq:elec_energy1},
\ref{eq:elec_energy2}, and Fig. \ref{fig:temporal}) and in the case
of narrow jets. Therefore, the results derived in Eq.
\ref{eq:B_cr_1} hold also in theses cases (mathematically, the only
difference is to replace the factor $2a_{jet}-1$ with a similar factor
$8 a_{jet}/3 -1$ when adiabatic energy losses are added). We can
therefore summarize that for $B_0 <B_{cr,1}$, the initial rapid
cooling is insignificant.

Our second acceleration model considers power law energy distribution of
the accelerated electrons above the peak of the Maxwellian and below a
maximum Lorentz factor $\gamma_{\max}$.
% with power law index $p$. 
Equations \ref{eq:elec_energy1} and \ref{eq:elec_energy2} that
describe the decay of the electrons Lorentz factor due to synchrotron
emission and adiabatic cooling are independent on the electrons
distribution, and hence are valid in this scenario as well.

The inclusion of a maximum Lorentz factor of the electrons energy
distribution $\gamma_{\max}$ implies that one can define an additional
critical value of the magnetic field, $B_{cr,0}$ in a similar way to
the definition of $B_{cr,1}$. It is defined as the value of the
magnetic field below which electrons at $\gamma_{\max}$ do not cool
significantly by synchrotron emission. Repeating the analysis carried
above shows that
\beq
\ba{lcl}
B_{cr,0} & = & 270 \, (2 a_{jet} -1)^{2/3} \nonumber \\
& & \times (\gamma_j/2)^{2/3} \,
(x_0/45 r_s)^{-2/3} \;  {\rm G}, \nonumber \\
\epsilon_{B,cr,0} & = & 10^{-8} \, (2 a_{jet} -1)^{4/3} \nonumber \\
& & \times
(\gamma_j/2)^{4/3} \,  (x_0/45 r_s)^{-4/3} {\bar u_0}^{-1} \, r_{0,1}^{2}. 
\ea
\label{eq:B_cr_0}
\eeq

We thus conclude that for very low values of the magnetic field, $B_0
< B_{cr,0}$, the distribution of the power law accelerated electrons
in the entire energy range is unaffected by the initial rapid cooling
phase.  If the magnetic field is stronger, $B_{cr,0} < B_0 <
B_{cr,1}$, then the highest end of the electrons distribution is
cooled rapidly close to the jet base, while for electrons at the low
end of the distribution, with $\gamma \gtrsim \gamma_{\min}$, the
initial cooling is inefficient. As a result, the main effect of the
initial rapid cooling is a shift of the highest energy tail in the
electrons energy distribution to lower energies. This result affects
the observed spectrum at high energies, and will be discussed in
\S\ref{sec:results2}.

For higher values of the magnetic field, the initial rapid cooling of
the energetic electrons is significant.  In the pure synchrotron case and
$a_{jet} > 1/2$, following the initial rapid decay, the electrons
Lorentz factor asymptotes to a constant value. Since in this case 
%of large magnetic field and wide jets, 
at $t \gg t_0$ the second term in
the denominators of both Eqs. \ref{eq:elec_energy1} and
\ref{eq:elec_energy2} are larger than unity, one finds a simple
analytical expression for the energetic electrons asymptotic momentum,
\beq
(\gamma \beta)_f = {\gamma \beta} (t \gg t_0) \simeq {6 \pi m_e
  c (2 a_{jet} - 1) \over \sigma_T B_0^2 t_0}.
\label{eq:theta_final}
\eeq 
The electrons Lorentz factor at the end of the rapid cooling phase is
therefore independent on its initial Lorentz factor, and is inversely
proportional to $B_0^2$ (see Fig. \ref{fig:temporal}). Similar to
the discussion that followed Eq. \ref{eq:B_cr_1}, we point
that while Eq. \ref{eq:theta_final} was derived for the pure
synchrotron case, a similar equation holds when adiabatic losses are
included.

If the magnetic field is in the range $B_{cr,0} < B_0 < B_{cr,1}$ then
the rapid cooling affects only electrons at the high end of the power
law energy distribution, while if the initial distribution is a
Maxwellian, the entire electrons distribution is unchanged. 
For stronger value of the magnetic field, $B_0 > B_{cr,1}$, electrons
at and above the peak of the Maxwellian rapidly cool. Since the 
electrons Lorentz factor at the end of the rapid cooling $\gamma_f$ is
independent on its initial Lorentz factor, then in both the Maxwellian
and power law acceleration scenarios we expect a quasi-Maxwellian
distribution to be formed at the end of the initial rapid cooling phase 
(see the numerical results in the figures below). This
quasi-Maxwellian distribution is characterized by temperature
$\theta_{el,f} = \gamma_f / 3$.

During the initial rapid decay phase, the magnetic field is nearly
constant. However, the characteristic frequencies $\nu_{peak}$ and
$\nu_{thick}$ depend on the electrons temperature, and therefore vary
as the electrons cool (see Eqs. \ref{eq:nu_p},
\ref{eq:nu_break}). While the peak frequency $\nu_{peak}$ decreases
when the electrons lose their energy, the break frequency
$\nu_{thick}$ increases.  Therefore, if the electrons cool to below a
critical temperature,
\beq
\theta_{cr} = \left({4 \over 3} \right)^{1/3} \left( {\pi^2 u_0 q r_0
    \over 9 B_0 m_p c^2} \right)^{1/5} = 30 \; {\bar u_0}^{1/10}
\eB^{-1/10},
\label{eq:theta_cr}
\eeq 
then at the end of the rapid cooling phase, $\nu_{thick,f} \equiv
\nu_{thick}(x=x_0 ; \theta_{el} = \theta_{el,f}) \geq
\nu_{peak,f}$. The condition that the electrons
cool to temperature $\theta_{el,f} \leq \theta_{cr}$ can be phrased as
a condition on the strength of the magnetic field at the base of the
jet. Using Eq. \ref{eq:theta_final}, one finds that this happens
for $B_0 \geq B_{cr,2}$, where
\beq
\ba{lcl}
B_{cr,2} & = & 8.0 \times 10^4 \, (2 a_{jet} -1)^{5/9} \nonumber \\
& & \times 
(\gamma_j/2)^{5/9} \, (x_0/45 r_s)^{-5/9} \, {\bar u_0}^{-1/9} \,
r_{0,1}^{1/9} \; {\rm G}, \nonumber \\
\epsilon_{B,cr,2} & = & 8.0 \times 10^{-4} \, (2 a_{jet} -1)^{10/9} \nonumber \\ 
& & \times 
(\gamma_j/2)^{10/9} \,  (x_0/45 r_s)^{-10/9} \, {\bar u_0}^{-11/9} \,
r_{0,1}^{20/9}. 
\ea
\label{eq:B_cr_2}
\eeq
We thus conclude that if the magnetic field is intermediate, $B_{cr,1}
< B_0 < B_{cr,2}$, then at the end of the initial cooling phase, still the
peak of the emission is in the optically thin region, $\nu_{peak,f} \geq
\nu_{thick,f}$. For higher values of the magnetic field, at the end of
the rapid cooling phase the peak of the emission is obscured,
$\nu_{thick,f} \geq \nu_{peak,f}$. 
%As we will show below, this result has
%a significant effect on the observed flux at the different wavebands.

\subsection{Asymptotic variations of the characteristic frequencies
  along the jet and the transition frequency}
\label{sec:parms2}

The change in the electrons energy distribution and the decay of the
magnetic field along the jet imply that the characteristic emission
frequencies, $\nu_{thick}$ and $\nu_{peak}$ vary along the
jet. Equations \ref{eq:elec_energy1} and \ref{eq:elec_energy2} give
the decay laws of the electrons energy along the jet, and thus enable
to calculate the variation of these frequencies. Although these
equations show complex dependence of the electrons energy in time, the
discussions in \S\ref{sec:elec_temporal_behavior} and
\S\ref{sec:B_critical} show that following an initial rapid decay
phase, the electrons decay laws at times $t \gtrsim t_0$ asymptote to
simple functions, that can be described analytically.  Therefore,
simple analytical description of the characteristic frequencies
temporal dependence is obtained in this limit.

If adiabatic energy losses are ignored, then for $a_{jet} > 1/2$ the
electrons temperature asymptotes to a constant, finite value. Thus, we
consider the asymptotic temporal dependence $\theta_{el} \propto t^0
\propto r^0$.  In this case, the electrons number and energy densities
drop as $n_{el, {\rm \, tot}} (r) \propto r^{-2}$, $u(r) \propto
r^{-2}$, resulting in a decay of the proportionality constant $A(r)
\propto r^{-2}$. Since the magnetic field also decays as $B(r) \propto
r^{-1}$, one finds from Eqs. \ref{eq:nu_p} and \ref{eq:nu_break}
that the characteristic frequencies decay in a similar way,
\beq
\nu_{peak} \propto \nu_{thick} \propto r^{-1} \propto x^{-a_{jet}}.
\label{eq:var1}
\eeq

When adiabatic energy losses are considered, for $a_{jet} > 3/8$, at
$t \gtrsim t_0$ the electrons temperature decays as $\theta_{el}(t)
\propto t^{-2 a_{jet}/3} \propto r^{-2/3}$. The electrons number
density drops as $n_{el, {\rm \, tot}}(r) \propto r^{-2}$, and
therefore the electrons energy density decays as $u(r) \propto
r^{-8/3}$. As a result, the characteristic frequencies decay according to
\beq
\ba{lcl}
\nu_{peak} & \propto & B \theta_{el}^2 \propto r^{-7/3} \propto x^{-7
  a_{jet}/3} \nonumber \\
\nu_{thick} & \propto &  (u^3 B^2 r^3)^{1/5} \theta_{el}^{-1} \propto
r^{-11/15} \simeq x^{-2 a_{jet}/3}. 
\ea
\label{eq:var2}
\eeq
The different $x$-dependent of the two frequencies imply that when
adiabatic energy losses are considered, even if at the base of the jet
$\nu_{peak,0} > \nu_{thick,0}$, then far enough along the jet, at
$x > x_{trans}$, $\nu_{thick} (x) > \nu_{peak}(x)$. In order to
estimate the transition distance $x_{trans}$, we discriminate between
two cases:  

(I) $B_0 < B_{cr,1}$: In weak magnetic field, the electrons do not
undergo the initial rapid cooling phase, and one finds that
$x_{trans}/x_0 = (\nu_{peak,0}/\nu_{thick,0})^{5/8 a_{jet}}$, or 
\beq
{x_{trans} \over x_0} = (250)^{1/a_{jet}} \; {\bar u_0}^{-3/(16 a_{jet})} \, 
\ee^{15/(8 a_{jet})} \, \eB^{3/(16 a_{jet})}.
\label{eq:x_transition}
\eeq
At $x_{trans}$, $\nu_{thick}(x_{trans}) = \nu_{peak}(x_{trans})$, and both are equal to
\beq
\nu_{trans}(B_0 < B_{cr,1}) = 3.7 \times 10^{11} \; {\bar u_0}^{15/16} \,
r_{0,1}^{-1}\, \ee^{-19/8} \, \eB^{1/16} {\rm \; Hz}.
\label{eq:nu_trans} 
\eeq
Thus, a change in the spectrum at $\nu_{trans}$ is expected.

(II) $B_{cr,1} < B_0$: In this case the electrons rapidly
cool before their temperature asymptotes.  Thus, at the end of the
rapid cooling, the ratio between the peak and break frequencies is
$\nu_{peak,f}/\nu_{thick,f} = \nu_{peak,0}/\nu_{thick,0} \times
(\theta_{el,f}/\theta_{el,0})^3$. Following the initial rapid cooling
the asymptotic behaviour of the two frequencies follows a similar
decay law to the one considered above in Eq.
\ref{eq:var2}, and therefore one finds that the transition from
optically thick to optically thin emission occurs at frequency
\beq
\ba{lcl}
\nu_{trans}(B_0 > B_{cr,1}) & = & 1.1 \times 10^{15} \;
 (8 a_{jet}/3 -1)^{-19/8} \nonumber \\
& & \times  {\bar u_0}^{53/16} \,
r_{0,1}^{-23/4}\, \, (\gamma_j/2)^{-19/8} \nonumber \\
& & \times (x_0/45 r_s)^{19/8}  \,
\eB^{39/16} {\rm \; Hz}. 
\ea
\label{eq:nu_trans2} 
\eeq 
By definition, if $B_0 > B_{cr,2}$ then the transition to the
optically thick emission occurs already at the base of the jet.

For narrow jets, $a_{jet}<1/2$, the electrons temperature decays as
$\theta_{el} \propto x^{2 a_{jet} -1}$, and their energy density drops
as $u \propto x^{-1}$. As a result, the peak frequency decays as
$\nu_{peak}(x) \propto x^{3 a_{jet} -2}$, while the break frequency as
$\nu_{thick} \propto x^{(-9 a_{jet} +2)/5}$. We find that the
transition frequency depends on the jet geometry in a complex way:
\beq
\ba{lcl}
\nu_{trans}(a_{jet} < 1/2) & \approx & 10^{17} \times (7 \times 10^3)^{A_1} 
\nonumber \\
& & \times
 {\bar u_0}^{1/2-3A_2/10} \,
r_{0,1}^{-1}\, \nonumber \\ & & \times \ee^{2+3A_2}  \,
\eB^{1/2+3A_2/10} {\rm \; Hz}, 
\label{eq:nu_trans3} 
\ea
\eeq
where  $A_1(a_{jet}) = (15 a_{jet} -10)/(12-24 a_{jet})$, and
$A_2(a_{jet}) = (12 - 24 a_{jet})(3 a_{jet}-2)/5$. The transition
frequency varies as a power in $A_1 (a_{jet})$.  For jet geometry
$a_{jet} \leq 1/3$, one finds that $\nu_{trans} \approx
10^{13}$~Hz, however $\nu_{trans}$ drops sharply for higher values of
$a_{jet}$: for $a_{jet} = 0.40$, $\nu_{trans} \approx 10^{10}$~Hz,
while for  $a_{jet} = 0.45$, $\nu_{trans} \approx 10^{6}$~Hz.

\subsection{Observed break frequencies from synchrotron emission along the
  jet}
\label{sec:frequencies2}

The analysis carried above implies that the observed flux from the
entire jet (as opposed to a single jet segment) is characterized by
several break frequencies. We present here a general discussion on the
nature of these break frequencies. This is done in order to establish
a general basis for the detailed discussion on the various models
(Maxwellian vs. power law, pure synchrotron vs. adiabatic, etc.)
discussed in the following sections. 

We are able to identify five observed break frequencies. However, some
of these frequencies are pronounced only under certain
conditions. These frequencies are:

(I) $\nu_{peak,0}$ (Eq. \ref{eq:nu_p}). This break frequency is
inherent to any model that contains a low energy Maxwellian
distribution, and is therefore expected in all the scenarios
considered.

(II) $\nu_{\max,0}$ (Eq. \ref{eq:nu_max}) is expected in models
in which a power law distribution of the accelerated electrons exist,
since any acceleration model necessarily have a high energy cutoff.

(III) When the magnetic field is higher than a minimum value, the
electrons at the high end of the distribution rapidly cool. 
%For a Maxwellian
%distribution of the electrons this minimum value is $B_{cr,1}$, while
%for power law distribution it is $B_{cr,0}$. 
As a result, % in this scenario 
the cooled electrons at the base of the jet (with Lorentz factor $\gamma_f$
given by Eq. \ref{eq:theta_final}) emit synchrotron radiation at
characteristic frequency $\nu_{f} = (3/4\pi) (q B_0 /m_e c) {\left(\gamma
  \beta \right)_f}^2$. In determining the observed
frequency, one needs to discriminate between the different cases.
For a Maxwellian distribution and $B_{cr,1} < B_0 < B_{cr,2}$,
$\nu_{peak,f} > \nu_{thick,f}$, and therefore 
emission at $\nu_{peak,f}$ is not obscured. Thus, in this case
\beq
\ba{lcl}
\nu_{fast} (B_0 \leq B_{cr,2}) & \equiv & \nu_{peak,f} =   {3 \theta_{el,f}^2
  \over 4 \pi} {q B_0 \over m_e c} \nonumber \\
& \simeq &
1.6 \times 10^{14} \; (2 a_{jet} -1)^{2} \, {\bar u_0}^{-3/2} \,
r_{0,1}^{3} \nonumber \\ & & \times \eB^{-3/2} 
\, (\gamma_j/2)^{2} \, (x_0/45 r_s)^{-2} 
{\rm Hz}.
\ea
\label{eq:nu_fast}
\eeq
Note that while Eq. \ref{eq:nu_fast} was derived for a Maxwellian
distribution, it holds for power law distribution as well as long as
$B_{cr,0} < B_0 < B_{cr,2}$, when $\theta_{el,f}$ is replaced by
$(\gamma \beta)_f$.

For stronger magnetic field, $B_0 > B_{cr,2}$, the electrons rapidly
cool to temperature below the critical temperature $\theta_{cr}$ (see
Eq. \ref{eq:theta_cr}). Once this happens, the peak of the
synchrotron emission is in the optically thick regime, and is
therefore obscured: $\nu_{peak,f} < \nu_{thick,f}$.  The observed
break in the flux therefore occurs at the transition frequency from
the optically thin to the optically thick emission. By definition,
this transition frequency occurs as the electrons cool to
$\theta_{cr}$.  Emission from electrons at this temperature peaks at
frequency 
\beq
\ba{lcl}
\nu_{fast}(B_0 \geq B_{cr,2}) & =  & {3 \theta_{el,cr}^2 \over 4 \pi} {q
  B_0 \over m_e c} \nonumber \\ & \simeq & 
4.0 \times 10^{14} \; {\bar u_0}^{7/10} \,
r_{0,1}^{-1} \, \eB^{3/10} {\rm \, Hz}.
\ea
\label{eq:nu_fast2}
\eeq

(IV) When adiabatic energy losses are included, or for narrow jets
$a_{jet} < 1/2$, the analysis in \S\ref{sec:parms2} shows that for
$B_0 < B_{cr,2}$ there exists a transition frequency $\nu_{trans}$,
whos value was calculated in Eqs. \ref{eq:nu_trans} --
\ref{eq:nu_trans3}. For high value of the magnetic field, $B_0 >
B_{cr,2}$, the emission peak is in the optically thick part of the
spectrum already during the initial rapid cooling. As a result, in
this case $\nu_{trans} = \nu_{fast}(B_0 \geq B_{cr,2})$, as is defined
in Eq. \ref{eq:nu_fast2}.

(V) The last transition frequency, which we denote here as $\nu_{low}$
is relevant in a scenario of power law energy injection, when
adiabatic energy losses (or narrow jets) are considered. Cooling of
the electrons at the highest end of the distribution implies that the
peak frequency of synchrotron emission from these electrons,
$\nu_{\max}(x)$ decays in a similar way to the decay of $\nu_{peak}$:
for wide jets with adiabatic energy losses, asymptotically
$\nu_{\max}(x) \propto x^{-7 a_{jet}/3}$. This decay is faster than
the decay of $\nu_{thick}$, and therefore at a given distance along
the jet $x_{low}$, these two frequencies become similar, and equal to
$\nu_{low}$. Since $\gamma_{\max} > \gamma_{\min}$, necessarily
$x_{low} \geq x_{trans}$, and hence $\nu_{low} \leq \nu_{trans}$. We
further note that if $B_0 > B_{cr,1}$, then at the end of the rapid
cooling, all the electrons with initial Lorentz factor above
$\gamma_{\min}$ cool to the same Lorentz factor, $\gamma_f$. As a
result, in this case $\nu_{low} = \nu_{trans}$.
 
We therefore concentrate in the regime $B_0 < B_{cr,1}$. 
Equation \ref{eq:var2} gives the decay law of $\nu_{thick}$ under the
assumption of Maxwellian energy distribution, and is therefore valid
as long as $\nu_{thick} < \nu_{peak}$, or equivalently $x <
x_{trans}$. At larger distances $x > x_{trans}$, the decay law of
$\nu_{thick}$ is modified and depends on the power law index $p$ of
the accelerated electrons.

In calculating the decay law of $\nu_{thick}$ in this case, we largely
follow the treatment by \citet{Kaiser06}. The electrons number density
between $\gamma_{\min}$ and $\gamma_{\max}$ can be written as $n_{el}
(\gamma) d\gamma = k(x) \gamma^{-p} d\gamma$. Conservation of
particles number along the jet can be written as $k(x) \gamma^{-p}
d\gamma = (\Delta V_0 / \Delta V) k(x_0) \gamma_0^{-p} d\gamma_0$,
where $\Delta V$ is the volume occupied by the electrons and the
subscript '0' refer to the values at $x=x_0$. Using Eq.
\ref{eq:ad1} for adiabatic energy losses along the jet and using
$\Delta V \propto r^2$, one finds that $k(x)=k(x_0) (r/r_0)^{-(2p+4)/3}$.

In this case of power law distribution, the emissivity is $j_{\nu}
\propto k B^{(p+1)/2} \nu^{-(p-1)/2}$ and the optical depth is
$\tau_{\nu} \propto r k B^{(p+2)/2} \nu^{-(p+4)/2}$
\citep[e.g.,][]{Rybicki79}. Using $\nu_{thick} = \nu|_{\tau_{\nu} =
  1}$, one finds
\beq
\nu_{thick} \propto \left( k r B^{(p+2)/2} \right)^{2/(p+4)} \propto
r^{-\hat A_1},
\label{eq:nu_break_last}
\eeq 
where $\hat A_1(p) = (8+7p)/3(p+4)$.
We emphasis that this decay law of $\nu_{thick}$ is relevant only for
$x > x_{trans}$, while at smaller radii, the analysis carried in
\S\ref{sec:parms2} holds. 

Repeating a similar analysis to the one carried in \S\ref{sec:parms2},
one finds that for $B_0 < B_{cr,0}$,  
\beq
\ba{lcl}
\nu_{low} & \approx & 2.6 \times 10^5 \;  {\bar u_0}^{(3 \hat A_2 /16)
  (5- \hat A_1)} \,
r_{0,1}^{-15 \hat A_2/16 }\nonumber \\
& & \times \ee^{(\hat A_2/8)(15 \hat A_1 - 19)} \,
\eB^{( \hat A_2/16)(2
  \hat A_1+1)} {\rm \; Hz}, \nonumber \\
& \approx  & 2.6 \times 10^5 \;  {\bar u_0}^{2.5} \, r_{0,1}^{-2} \,
\ee^{-10} \, \eB^{-0.4} {\rm \; Hz}, 
\ea
\label{eq:nu_low}
\eeq
where $\hat A_2 (p) = (7/3)/[(7/3)+\hat A_1(p)]$, and the second line gives the
approximate dependence for $p$ in the range  $2.0 \leq p \leq 2.5$.

Similarly, for $B_{cr,0} < B_0 < B_{cr,1}$,
\beq
\nu_{low} \approx  10^{12} \; {\bar u_0}^{4.25} \,
r_{0,1}^{-5.5}\, \, (\gamma_j/2)^{-2} \, (x_0/45 r_s)^{2}  \,
\epsilon_{B,-5}^{1.5} {\rm \; Hz}. 
\label{eq:nu_low2}
\eeq

Calculation of $\nu_{low}$ for narrow jets, $a_{jet} < 1/2$ is
straightforward, yet cumbersome. For $2.0 \leq p \leq 2.5$ and
$a_{jet} \lesssim 0.1$, we find $\nu_{low} \approx 10^{12}$~Hz, while
for wider jets, $\nu_{low}$ rapidly decays: $\nu_{low} \approx
10^{10}$~Hz for $a_{jet} = 0.2$, and $\nu_{low} \approx 10^{7}$~Hz for
$a_{jet} = 1/3$. As we will show in \S\ref{sec:results5}, for
narrow jets it is impossible to obtain a flat spectrum at the radio
band, as suggested by observations.  Therefore, we will give only
brief explanation on the possible spectrum obtained in this scenario.

We further discuss the role of these five transition frequencies on
the observed spectrum in the following sections, when we present a
detailed analysis of the various spectras that can be obtained. The
transition frequencies are marked in Figg. \ref{fig:2} --
\ref{fig:9} when discussing the various possibilities.

\section{Emission from initial Maxwellian distribution of electrons
  without adiabatic cooling}  
\label{sec:results1}

In this section we present the results of a model that includes only
synchrotron cooling of the electrons, i.e., we neglect adiabatic
energy losses. While adiabatic energy losses occur as the gas expands,
both the formation and confinement mechanisms of jets are still not
fully understood. It could thus be that the expansion of particles
inside the jet is not purely adiabatic, and that the electrons are
heated as they propagate along the jet (e.g., by lateral shock waves).
Neglection of the adiabatic energy losses can thus be considered as an
extreme, yet still physically plausible scenario.

We further concentrate in this section in a scenario in which the
acceleration process results in a Maxwellian distribution of the
energetic electrons. This assumption has both theoretical motivation,
and is used for illustrative purpose.  From a theoretical perspective,
as discussed in \S\ref{sec:intro} the mechanism that accelerates
particles to high energies (e.g., via shock waves) is not understood
from first principles. While it is likely that some fraction of the
electrons are accelerated to high energy power law tail above the peak
of the Maxwellian, the fraction of electrons in the energetic tail is
uncertain. Thus, we consider this scenario as being a physically
plausible scenario of electrons acceleration process.  In addition,
studying this scenario is helpful in understanding the spectral
dependence on the uncertain value of the magnetic field and possible
jet geometries. We expand our model in \S\ref{sec:results2} to the
cases of power law distribution of the energetic electrons and to
consider adiabatic energy losses. 

Here and in sections \S\ref{sec:results2} -- \S\ref{sec:results4} we
focus on scenarios of wide jets. Narrow jet scenario is discussed in
\S\ref{sec:results5}.

\subsection{Weak magnetic field: production of flat radio spectra in
  conical jets}
\label{sec:results1a}

For weak magnetic field at the jet base, $B_0 < B_{cr,1}$, and for  
wide jets, $a_{jet} > 1/2$, 
% the initial
%rapid cooling is insignificant (see Eq. \ref{eq:B_cr_1}).  Since
%following this initial cooling, for wide jets, $a_{jet} > 1/2$, the
%electrons do not cool further as they propagate along the jet, we
%conclude that in this scenario 
electrons cooling is insignificant in
the entire jet (this scenario is illustrated in figure
\ref{fig:temporal}, $B_0 = 10^4$~G).  The insignificance of electrons
cooling implies that the electrons temperature $\theta_{el}$ is
constant along the jet, $\theta_{el}(x) \simeq \theta_{el,0}$. The
peak and break frequencies thus decay in accordance to Eq.
\ref{eq:var1}, $\nu_{peak} \propto \nu_{thick} \propto x^{-a_{jet}}$.

\begin{figure}
\plotone{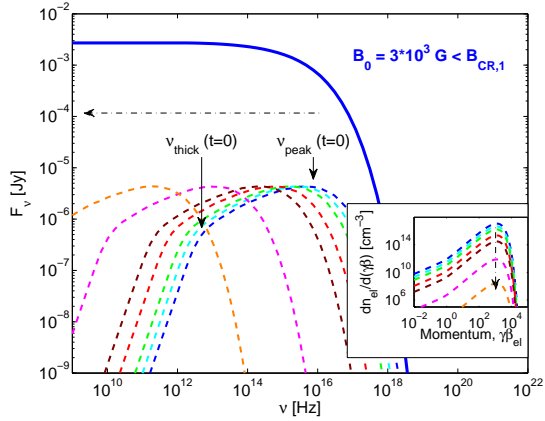}
\caption{Example of the spectra obtained from an initial Maxwellian
  distribution of electrons radiating in weak magnetic
  field, $B_0 = 3\times 10^3{\rm \, G} < B_{cr,1}$, when adiabatic
  energy losses are neglected. The values of the
  free model parameters are the same as the ``canonical'' values
  taken in \S\ref{sec:2.1.1}. Thus, we consider flow parameters
  resulting in ${\bar u_0} = 1$, $x_0 = 45 r_s$, $r_0 = 10 r_s$, and
  $\ee = 1$. We further consider a conical jet, $a_{jet} = 1$. As the
  distance to the object we take the distance to XTE J1118+480, $d =
  1.8$~kpc. The inner panel shows the electrons energy distribution at
  increasing distances along the jet. As the electrons temperature is
  nearly constant, their density decays as $n_{el,tot} \propto
  r^{-2}$. In the main panel we show the resulting flux (solid line),
  and its decomposition into the flux emitted from the different jet
  segments (dash lines). Marked in arrows are the characteristic
  frequencies $\nu_{peak,0}$ and $\nu_{thick,0}$. These frequencies
  are clearly seen when looking at emission from an arbitrary jet
  segments at $x>x_0$, however the overall spectra below
  $\nu_{peak,0}$ is flat, $F_\nu \propto \nu^0$ (for $a_{jet}=1$, see
  text). The dash-dotted arrows show the direction of the evolution of
  the electrons distribution and the emitted flux at different times,
  which are equivalent to distance along the jet.}
\label{fig:2}
\end{figure}

The observed flux at frequency $\nu \leq \nu_{peak,0}$ can be
calculated by integrating the emitted flux from the different jet
segments (this is illustrated in Fig. \ref{fig:2}). In order to carry
an analytical calculation, we approximate emission from a jet segment
at position $x.. x+dx$ as contributing only at frequencies $\nu \leq
\nu_{peak}(x)$ (at higher frequencies the flux decays exponentially).
The decay of the magnetic field implies that the peak frequency
$\nu_{peak}(x)$ decays along the jet. Therefore, at a given frequency
$\nu \leq \nu_{peak,0}$, the contribution to the flux is from jet
regions only up to a position $x_{\max}$ for which
$\nu_{peak}(x_{\max}) = \nu$.  The decay law of the magnetic field
thus gives $x_{\max} = x_0 [\nu/\nu_{peak,0}]^{-1/a_{jet}}$. Using the
contribution to the flux from the optically thin regions of the
spectrum (Eq. \ref{eq:F_nu_thin}) one obtains
\beq
\ba{lcl}
F_\nu & \simeq & {2 r_0^2 A_0 \over  9 d^2} {\theta_{el_0}^3 q^3 B_0
  \over m_e c^2} \nonumber \\
& & \times 
\int_{x_{\min}}^{x_{\max}} \left({x \over x_0}\right)^{-a_{jet}} 
\left({3 \nu \over 4 \nu_{peak,0}}\right)^{1/3} 
\left({x \over x_0}\right)^{a_{jet}/3} dx  \nonumber \\
& = & \left({3 \over 4}\right)^{1/3} 
{2 r_0^2 \over  9 d^2}{u_0 q^3 B_0 x_0 \over (1-2a_{jet}/3)
  m_e c^2 m_p c^2} \left({ \nu \over \nu_{peak,0}}
\right)^{1-1/a_{jet}}.  
\ea
\label{eq:F_weak}
\eeq
The lower integration boundary is taken here as $x_{\min} = x_0$,
however the exact boundary is irrelevant as long as $x_{\max} \gg
x_{\min}$. We thus conclude that for weak magnetic field, the observed
flux at $\nu < \nu_{peak,0}$ depends on the jet geometry via $F_\nu
\propto \nu^{1-1/a_{jet}}$ (see Fig. \ref{fig:2}).

For conical jet, $a_{jet} = 1$, we therefore find that the flux below
$\nu_{peak,0}$ is constant (i.e., a flat spectrum, $F_\nu \propto
\nu^0$). While this result is similar to the result first obtained by
\citet{BK79}, here it has a different physical origin. In
\citet{BK79} the origin of the flat radio spectrum is the decay of the
break frequency $\nu_{thick}$ along the jet, while here it is the
decay of peak frequency, $\nu_{peak}$. The decay law of the break
frequency is similar to that of the peak frequency, and thus its
existence cannot be observed when the integrated flux along the jet is
considered.

We further point that in this regime of $B_0 < B_{cr,1}$, the total
(integrated) observed flux below $\nu_{peak,0}$ linearly depends on
$B_0$, and on the position of the jet base, $x_0$. We show in figure
\ref{fig:2} the results of an exact numerical calculation of the
observed flux and its decomposition to emission from different jet
segments obtained in this case.  The details of the numerical code are
given in appendix \ref{sec:numerics}.

\subsection{Intermediate magnetic field: spectral slope $F_{\nu}
  \propto \nu^{-1/2}$ at the  Optic/UV bands}
\label{sec:results1b}

\begin{figure}
\plotone{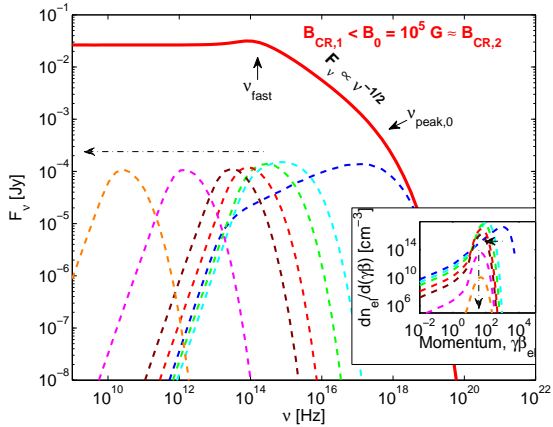}
\caption{Example of the spectra obtained from an initial Maxwellian
  distribution of electrons radiating in an intermediate
  magnetic field, $B_{cr,1} < B_0 = 10^5{\rm \, G} \approx B_{cr,2}$,
  when adiabatic energy losses are neglected. Apart from the magnetic
  field, all the other parameters values are the same as in
  Fig. \ref{fig:2}. Inner panel: electrons energy distribution at
  different regions in the jet. The dash-dotted arrows show the
  temporal evolution: at first the electrons rapidly cool close to the
  jet base, while maintaining an approximate Maxwellian
  distribution. At a later stage, the electrons number density
  decreases as $n_{el,tot} \propto r^{-2}$. Main panel: the resulting
  spectrum (solid line) and a decomposition to the spectrum emitted
  from different segments (dash lines). We mark the transition
  frequencies $\nu_{peak,0}$ and $\nu_{fast}$. The initial rapid
  electrons cooling results in a fast decay of $\nu_{peak}(t)$, which
  leads to a flux decay $F_\nu \propto \nu^{-1/2}$ at the range
  $\nu_{fast} < \nu < \nu_{peak,0}$. Following the rapid decay, the
  two characteristic frequencies $\nu_{thick}$ and $\nu_{peak}$ have
  nearly the same values (see text) and evolve in a similar way along
  the jet, as is seen by the decomposition. As a result, below
  $\nu_{fast}$, the flux decays in a similar way to the decay in the
  weak magnetic field scenario, described in Fig. \ref{fig:2}.  }
\label{fig:3}
\end{figure}

In this section we consider intermediate magnetic field at the jet
base, $B_{cr,1} < B_0 < B_{cr,2}$. By definition, for $B_0 > B_{cr,1}$
the electrons rapidly cool close to the jet base (see
\S\ref{sec:B_critical}, Eqs. \ref{eq:B_cr_1}), while the
requirement $B_0 < B_{cr,2}$ implies that at the end of the rapid
cooling phase, still $\nu_{thick} < \nu_{peak}$ (see Eq.
\ref{eq:B_cr_2}). The rapid cooling introduces an additional break
frequency in the observed spectrum, $\nu_{fast}$ (see
\S\ref{sec:frequencies2} Eq. \ref{eq:nu_fast}, and figure
\ref{fig:3}). 

As explained in \S\ref{sec:elec_temporal_behavior} and
\S\ref{sec:frequencies2} in this case of initial rapid cooling, the
analytical calculations are carried in two steps. We first calculate
the emissivity during the rapid cooling phase, under the assumption
that the magnetic field is constant during this phase, and is equal to
$B_0$. Emission at this phase contributes to the flux at the frequency
range $\nu_{fast} \leq \nu \leq \nu_{peak,0}$. At a second step, we
calculate the emissivity during the rest of the electrons propagation
along the jet, assuming that their temperature is constant in this
phase (given by Eq. \ref{eq:theta_final}), and that only the
magnetic field decays. At this stage, the electrons emission
contribute to the flux at lower frequencies $\nu < \nu_{fast}$. The
results of the full numerical calculations presented in figure
\ref{fig:3}, confirms the validity of the analytical approximations.

The flux at $\nu_{peak,0}$ is
estimated by assuming that the entire electrons distribution is
concentrated at a single energy (characterized by Lorentz factor
$\gamma_{\min}$) and that the emission in monochromatic, at
$\nu_{peak,0}$. Thus, the total power emitted at $\nu_{peak,0}$ is
$P_\nu|_{\nu = \nu_{peak,0}} \simeq P_0/\nu_{peak,0} = 16\pi q^3
B_0/27 m_e c^2 \Lambda$. Here, $P_0 = 4 q^4 B_0^2 {(\gamma
  \beta)_{\min}}^2/9 m_e^3 c^5$ is the total power emitted, and the
factor $\Lambda \approx 10$ is introduced here to account for the
distribution in the electrons energy and for the fact that the
emission is not monochromatic.  The emissivity is $j_\nu|_{\nu =
  \nu_{peak,0}} = n_{el,tot} \times P_\nu|_{\nu = \nu_{peak,0}}/4\pi$
where $n_{el,tot} \approx u_0/m_p c^2$ is the number density of
electrons at the base of the jet. The observed flux from a segment of
length $\Delta x$ close to the base of the jet is therefore
\beq
\Delta F_\nu|_{\nu = \nu_{peak,0}} = {r_0^2 \over 2 d^2} {4 q^3 B_0
    \over 27 m_e c^2 \Lambda} {u_0 \over m_p c^2} \Delta x.
\label{eq:F_tot_1}
\eeq   
Comparison with Eq. \ref{eq:F_weak} derived in
\S\ref{sec:results1a} indeed shows that the results are similar
up to a numerical factor. However, the estimate here is based on
robust arguments, that can be (and will be) generalized in studying
the flux in the power law scenario discussed below.  

The length $\Delta x$ at which electrons radiation contributes to the
flux at $\nu_{peak,0}$ can be estimated by the characteristic time it
takes the electrons to lose their energy, Eq.
\ref{eq:elec_energy2}.  Since the energy loss is very rapid and occurs
within time $t \gtrsim t_0$, one can approximate the energy loss time
in the following way.  Defining $\epsilon = t/t_0 -1 = x/x_0 - 1$, one
finds that for $\epsilon \ll 1$, Eq. \ref{eq:elec_energy2} can be
written in the form
\beq
\gamma \beta(t \gtrsim t_0) \simeq \frac{\gamma \beta(t_0)}{1 + 
    {4 \sigma_T \over 3 m_e c} {B_0^2 \over 8 \pi} 
    \gamma \beta (t_0) t_0 \epsilon} \approx {6 \pi m_e c \over
    \sigma_T B_0^2 t_0 \epsilon},
\label{eq:elec_energy3}
\eeq
where the last equality holds for $\epsilon \gg 6 \pi m_e c/\sigma_T
B_0^2 \gamma \beta(t_0) t_0$.\footnote{Note that for $B_0 > B_{cr,1}$,
  by definition $6 \pi m_e c/\sigma_T
B_0^2 \gamma \beta(t_0) t_0 < 1$, and thus the solution is valid.}
The characteristic energy loss time of electrons with initial
momentum $\gamma\beta(t_0)$ can therefore be written as
\beq 
\Delta t = t_0 \epsilon \approx { 6 \pi m_e c \over \sigma_T B_0^2
  \gamma \beta(t_0)}.
\label{eq:t_loss} 
\eeq
Writing $\Delta x = \gamma_j \beta_j c \Delta t$ and using this result
in Eq. \ref{eq:F_tot_1}, one finds that the
flux at $\nu_{peak,0}$ is 
\beq
F_\nu|_{\nu = \nu_{peak,0}} = {4 q^3 q_j {\dot M}_{\rm disk} 
    \over 27 d^2 m_p c \sigma_T B_0 \theta_{el,0} \Lambda}.
\label{eq:F_nu_peak}
\eeq

At frequencies below $\nu_{peak,0}$ and above $\nu_{fast}$, the flux
can be calculated as follows. At this frequency range, the flux is
dominated by emission from electrons during their initial rapid
cooling phase.  As the electrons cool, the peak of the synchrotron
emission $\nu = \nu_{peak}(t)$ decays. The power emitted at $\nu$,
$P_\nu|_{\nu_{peak}(t)}$ is independent on the electrons energy (see
discussion and Eq. \ref{eq:F_tot_1}; note that since
$B=B_0$ is constant, Eq. \ref{eq:F_tot_1} is in fact valid for
emission at all the frequencies in range $\nu_{fast} \leq \nu \leq
\nu_{peak,0}$, when the appropriate length $\Delta x$ is taken).
Therefore, the observed flux at these frequencies $F_{\nu} (\nu_{fast}
< \nu<\nu_{peak,0})$ depends only on the electrons cooling time, or
cooling length $\Delta x$, during which synchrotron emission
contributes to the flux at frequency $\nu$. Using the decay law of the
electrons energy at early times derived above, Eq.
\ref{eq:elec_energy3}, one finds $\gamma \beta(t) \propto (t_0
\epsilon)^{-1} \propto \Delta x^{-1}$.  The peak emission frequency
depends on the electrons momentum via $\nu = \nu_{peak}(t) \propto B
(\gamma \beta)^2(t) \propto \Delta x^{-2}$. We therefore conclude that
$\nu(t) \propto \Delta x^{-2}$. Using again Eq.
\ref{eq:F_tot_1}, one immediately concludes that since $F_\nu \propto
\Delta x$, then for $\nu_{fast} \leq \nu \leq \nu_{peak,0}$ the flux
decays as $F_\nu \propto \nu^{-1/2}$. Note that this result is
independent on the jet geometry.

Once the electrons cool to their asymptotic temperature,
$\theta_{el,f}$, they continue to propagate along the jet without
further significant energy losses.  For intermediate value of the
magnetic field, $B_{cr,1} \leq B_0 \leq B_{cr,2}$, in spite of their
cooling still $\nu_{thick} < \nu_{peak}$. As a result, at low
frequencies $\nu < \nu_{fast}$, the analysis of the flux carried in
\S\ref{sec:results1a} holds. In particular, we conclude that for $\nu
< \nu_{fast}$, $F_\nu \propto \nu^{1-1/a_{jet}}$, and that the flux at
these frequencies linearly depends on $B_0$ and $x_0$.\footnote{The
  linear dependence of the flux below $\nu_{fast}$ on $B_0$ can be
  derived indirectly, from the analysis carried above. It is an
  immediate consequence of the facts that $\nu_{peak,0} \propto B_0$,
  $F_{\nu}|_{\nu_{peak,0}} \propto B_0^{-1}$, $\nu_{fast} \propto
  B_0^{-3}$ and $F_{\nu}|_{\nu_{fast} < \nu < \nu_{peak,0}} \propto
  \nu^{-1/2}$.}

Examples of the spectrum are presented in Fig. \ref{fig:3}. In the
inner inset we show the electrons energy distribution in different
regions along the jet (the arrows indicate the direction of the
temporal or equivalently the spatial evolution). In spite of the
early, rapid cooling, still the electrons maintain a quasi-Maxwellian
distribution along the jet, with asymptotic temperature given by
Eq. \ref{eq:theta_final}. The main panel shows the observed flux,
and its decomposition into the flux emitted from the different
segments along the jet. The flux decay $F_\nu \propto \nu^{-1/2}$ at
the range $\nu_{fast} < \nu < \nu_{peak,0}$ is pronounced. As can be
seen by the decomposition, contribution to the flux at these
frequencies is only from the innermost parts of the jet.

\subsection{Strong magnetic field: suppression of the flux at radio frequencies}
\label{sec:results1c}

If the magnetic field at the jet base is higher than $B_{cr,2}$, then
the electrons rapidly cool to temperature below the critical
temperature $\theta_{cr}$ (see Eq. \ref{eq:theta_cr}). Once this
happens, the peak of the synchrotron emission is in the optically
thick region, and is therefore obscured, $\nu_{peak} < \nu_{thick}$.
The observed break frequency $\nu_{fast}$ in this case is given by
emission from electrons at $\theta_{cr}$ and was calculated in
\S\ref{sec:frequencies2}, Eq. \ref{eq:nu_fast2}.

Emission at lower frequencies that occur as the electrons continuously
cool below $\theta_{cr}$ is obscured, since it is in regions of
high optical depth ($\nu_{peak} < \nu_{thick}$). As a result, for $B_0
> B_{cr,2}$ the peak of the observed flux is at frequency $\nu_{fast}$
given by Eq. \ref{eq:nu_fast2}, while at lower frequencies the
flux decays.  We therefore conclude that further increase of the
magnetic field above $B_{cr,2}$ results in an increase of the observed
spectral break frequency, $\nu_{fast}$, in accordance to the
dependence given in Eq. \ref{eq:nu_fast2} (as opposed to the
result obtained for $B_{cr,1} \leq B_0 \leq B_{cr,2}$, see equation
\ref{eq:nu_fast}).

Similar to the scenario presented in \S\ref{sec:results1b}, in
strong magnetic field the electrons lose most of their energy
rapidly. Thus, the calculation of the flux at $\nu_{peak,0}$ in 
Eq. \ref{eq:F_nu_peak} holds. Similarly, below $\nu_{peak,0}$ and
above $\nu_{fast}$, the flux decays as $F_{\nu} \propto \nu^{-1/2}$.
 
At lower frequencies, radio - IR, emission is dominated by electrons
propagating along the jet with temperature $\theta_{el,f} <
\theta_{cr}$.  Emission from these electrons is characterized by
$\nu_{peak}(x) < \nu_{thick}(x)$. For a Maxwellian distribution of
electrons, above $\nu_{peak}$ the flux decays exponentially. Thus, an
exact calculation of $\nu_{thick}$ requires solving a transcendental
equation. Using the results of \citet{JH79} for $z \gg 1$ (see
discussion below Eq. \ref{eq:nu_p}), and the optical depth given
in Eq. \ref{eq:tau}, the break frequency $\nu_{thick}$ can be
calculated by solving the Eq. $[3 \nu_{thick}(x)/ 2
\nu_{peak}(x)]^{1/3} = (2/3) \log[\pi^2 r(x) A q^2 c / 4
\nu_{thick}(x)]$. Here, we approximate the logarithm on the right hand
side as constant, which enables us to write $\nu_{thick}(x) = \alpha
\nu_{peak}(x)$. The parameter $\alpha$ depends on $B_0$.  We find
numerically that $\alpha$ has characteristic value of few tens
($\alpha \approx 3$ for $\epsilon_B \simeq \epsilon_{B,cr,2}$, $\alpha
\approx 60$ for $\epsilon_B = 10 \epsilon_{B,cr,2}$ and $\alpha
\approx 100$ for $\epsilon_B = 100 \epsilon_{B,cr,2}$).

With these approximation, one can calculate the expected flux at the
radio frequencies by integrating the flux in the optically thick part
of the spectrum (Eq. \ref{eq:F_nu_thick}) over the different jet
segments, in a similar way to the calculation of the radio flux
carried in \S\ref{sec:results1a}. Since the electrons cool to their
terminal temperature $\theta_{el,f}$ close to the jet base, at a given
frequency $\nu \leq \nu_{thick}(x)$ emission is obtained from jet
regions $x_0 \leq x \leq x_{\max}$, where $x_{\max} = x_0 (\nu/\alpha
\nu_{peak,f})^{-1/a_{jet}}$.
Integrating the flux in the optically thick part of the spectrum,
using Eq. \ref{eq:F_nu_thick} one obtains
\beq
\ba{lcl}
F_\nu & \simeq & {r_0 \over d^2} m_e \theta_{el,f} \nu^2 
\int_{x_{\min}}^{x_{\max}} \left({x \over x_0}\right)^{a_{jet}} dx  \nonumber \\
& = &
{r_0 \over d^2}{m_e \theta_{el,f} \nu_{peak,f}^2 x_0 \alpha^{1+1/a_{jet}}
\left({ \nu \over \nu_{peak,f}}\right)^{1-1/a_{jet}}}.  
\ea
\label{eq:F_strong}
\eeq

We therefore find that for $B_0 > B_{cr,2}$, the radio flux {\it
  decreases} with the increase of $B_0$, as $F_\nu \propto
\theta_{el,f}^5 B_0^2 \propto B_0^{-8} (\propto \epsilon_B^{-4})$. We
find numerically that the decay law is somewhat weaker, due to the
non-linear dependence of $\alpha$ on $B_0$.  We further find that the
radio flux depends on the jet size as $F_\nu \propto x_0^{-4}$, and on
the jet geometry in a similar way as in the weak magnetic field
scenario, $F_\nu \propto \nu^{1-1/a_{jet}}$.

\begin{figure}
\plotone{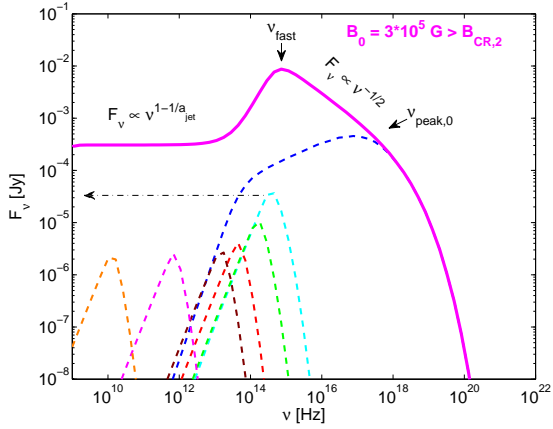}
\caption{Example of the spectra obtained from an initial Maxwellian
  distribution of electrons radiating in strong magnetic
  field, $B_0 = 3 \times 10^5{\, \rm{G}} > B_{cr,2}$, when adiabatic
  energy losses are neglected. All the other parameters are the same as
  in Fig. \ref{fig:2}. Clearly, the flux at high frequencies, $\nu >
  \nu_{fast}$ is similar to the case of lower magnetic field (see
  Fig. \ref{fig:3}), while at low frequencies, below $\nu_{fast}$ the
  flux is strongly suppressed. The decomposition of the flux to
  emission from the different jet segments, reveals that the origin of
  this suppression is the suppression of emission from the different
  segments, which are in the optically thick regime following the
  initial rapid cooling. The dash-dotted arrow shows the temporal
  evolution of the emitted flux along the jet.  Clearly, emission at
  high frequencies is from electrons during their initial fast cooling
  phase, and therefore the only effect of the strong magnetic field is
  a shift in $\nu_{fast}$ (see text for details). The electrons
  temporal evolution is very similar to the one presented by the inner
  panel in figure \ref{fig:3}, and is thus omitted.}
\label{fig:4}
\end{figure}

Example of the spectra obtained is presented in figure
\ref{fig:4}. Clearly, the flux at high frequencies, above $\nu_{fast}$
is similar to the flux obtained for weaker magnetic field (see figure
\ref{fig:3}). However, at lower frequencies, mainly at the radio band,
the flux is strongly suppressed due to the self absorption. This
result can naturally explain differences in radio emission that are
not accompanied by similar differences at higher bands in various
sources, by an adjustment of a single parameter- the magnetic field.

In Fig. \ref{fig:5} we compare the resulting spectra for different
values of the magnetic field. While this figure is derived for power
law distribution of the energetic electrons (see
\S\ref{sec:results2}), the resulting flux below $\nu_{peak,0}$ are
independent on the distribution of the electrons above $\gamma_{\min}$
and are thus similar for both cases. This figure illustrates how an
increase of the magnetic field from low values first leads to an
increases in the radio flux, which changes to a decrease in the
observed flux when $B_0 > B_{cr,2}$. The flux at the X band on the
other hand is much less sensitive to the value of the magnetic
field. We further discuss the implications of this in
\S\ref{sec:summary}.

\section{Emission from power law distribution of electrons without
  adiabatic cooling}
\label{sec:results2}

We next consider an initial distribution of the accelerated electrons
in the form of a power law energy distribution above the peak of the
Maxwellian and below a maximum Lorentz factor $\gamma_{\max}$, with
power law index $p$.  Equation \ref{eq:elec_energy2} that describes
the decay of the electrons momentum due to synchrotron emission is
independent on the electrons distribution, and hence is valid in this
scenario as well. As a result, most of the analysis carried in
\S\ref{sec:results1} is valid here. However, the inclusion of a power
law implies that there is an additional critical value for the
magnetic field, $B_{cr,0}$ (see \S\ref{sec:B_critical}, equation
\ref{eq:B_cr_0}). Thus, in calculating the resulting spectrum one
needs to discriminate here also between the cases $B_0 > (<)
B_{cr,0}$, since these two cases are qualitatively different. For weak
magnetic field, $B_0 < B_{cr,0}$ the electrons distribution in
unaffected by synchrotron cooling. In contrast, for $B_0 >
B_{cr,0}$, the high energy tail of the electrons distribution is
cooled significantly, while the low energy tail may be less
affected. As a result, the initial electrons energy distribution is
modified.

\subsection{Weak magnetic field}
\label{sec:results2a}

We first consider a very weak magnetic field at the jet base, $B_0 <
B_{cr,0}$.  In this case, electrons cooling in the entire energy range
is insignificant. Thus, standard synchrotron theory
\citep[e.g.,][]{Rybicki79} implies that above $\nu_{peak}$ and below
$\nu_{\max}$, the power emitted by a jet segment is $P_\nu \propto
\nu^{-(p-1)/2}$. The observed flux is calculated in a very similar way
to the calculation of the flux done in \S\ref{sec:results1}, by
integrating the contribution to the flux from the different segments
along the jet.  In particular, at low frequencies $\nu <
\nu_{peak,0}$, a similar analysis to the one carried in
\S\ref{sec:results1a} holds, and thus $F_\nu \propto B_0
(\nu/\nu_{peak,0})^{1-1/a_{jet}}$. This is illustrated in figure
\ref{fig:5}.

The arguments that led to Eq. \ref{eq:F_tot_1} can be generalized
to emission at high frequencies, $\nu > \nu_{peak}$ from an arbitrary
jet segment at position $x .. x + dx$, as long as electrons in this
jet segment contribute to the emission at this frequency. 
One therefore finds that the contribution
to the flux at high frequencies from a jet segment is given by
\beq
dF_\nu|_{\nu \geq \nu_{peak}} = {2 q^3 B
    \over 27 \pi d^2 m_e c^2 \Lambda} {q_j {\dot M}_{\rm disk} \over
      \gamma_j \beta_j m_p c} \left( {\nu \over \nu_{peak}}
    \right)^{-(p-1)/2} dx.
\label{eq:dF_nu_2}
\eeq  
In order to calculate the total flux, one needs to integrate over the
jet segments that contribute to the flux at a given frequency $\nu$.
Since in this case of weak magnetic field the electrons energy is
constant along the jet, the peak frequency decay is attributed only to
the decay of the magnetic field, and $\nu_{peak} = \nu_{peak,0}
(x/x_0)^{-a_{jet}}$.  The observed flux at frequency $\nu \geq
\nu_{peak,0}$ is therefore calculated by integrating the emission from
jet regions $x_0 \leq x \leq x_{\max}$,
\beq
\ba{lcl}
F_\nu|_{\nu \geq \nu_{peak,0}} & = & {2 q^3 B_0 \over 27 \pi d^2 m_e
  c^2 \Lambda} {q_j {\dot M}_{\rm disk} \over 
      \gamma_j \beta_j m_p c} \left( {\nu \over \nu_{peak,0}} 
    \right)^{-{(p-1) \over 2}} \nonumber \\
& & \times  {x_0 \over a_{jet}{(p+1) \over 2} -1}
    \left[ 1 - \left({x_0 
        \over x_{\max}}\right)^{a_{jet}{(p+1) \over 2} -1}\right].
\ea
\label{eq:F_tot_4}
\eeq  
For a given frequency $\nu \ll \nu_{\max}$, contribution to the flux
is from regions up to $x_{\max}(\nu)/x_0 =
(\nu/\nu_{\max})^{-1/a_{jet}} \gg 1$. The observed flux at high
frequencies thus depend on the jet geometry. For wide jets, $a_{jet} >
2/(p+1)$, the second term in the parenthesis of equation
\ref{eq:F_tot_4} can be neglected.  The flux above $\nu_{peak,0}$ is
therefore proportional to $F_\nu \propto B_0^{(p+1)/2}
\nu^{-(p-1)/2}$.  For narrower jets, $1/2 < a_{jet} < 2/(p+1)$ and
$\nu \ll \nu_{\max}$, the second term in the parenthesis of equation
\ref{eq:F_tot_4} dominates, and one finds $F_\nu \propto B_0^{(p+1)/2}
\nu^{1-1/a_{jet}}$. The case of $a_{jet} < 1/2$ is treated in
\S\ref{sec:results5}.

\subsection{Intermediate magnetic field: gradual change of the
  spectral slope at the X band}
\label{sec:results2b}

If the magnetic field is in the range $B_{cr,0} < B_0 < B_{cr,1}$ then
electrons at $\gamma_{\max}$ cool significantly at $t \gg t_0$, while
electrons at the lower end of the energy distribution, $\gamma \gtrsim
\gamma_{\min}$ do not significantly cool. This introduces a break at
$\nu_{fast}$, given by Eq. \ref{eq:nu_fast}. We note that for
$B_0 < B_{cr,1}$, $\nu_{fast} > \nu_{peak,0}$.  At the range
$\nu_{peak,0} < \nu < \nu_{fast}$ the analysis carried in
\S\ref{sec:results2a} is valid. Hence, for wide jets $a_{jet} >
2/(p+1)$, $F_\nu \propto \nu^{(p-1)/2}$. Similarly, at lower
frequencies, $\nu < \nu_{peak,0}$, $F_\nu \propto \nu^{1-1/a_{jet}}$.

At higher frequencies, $\nu > \nu_{fast}$, the spectrum is
modified. Since the energetic electrons rapidly cool, emission at
these high frequencies occurs only close to the jet base, where the
magnetic field is constant. The flux at these frequencies can
therefore be estimated by integrating the flux emitted from a jet
segment, given by Eq. \ref{eq:dF_nu_2}, while keeping $B=B_0$ and
$\nu_{peak} = \nu_{peak,0}$.

The length $\Delta x \equiv x_{\max}(\nu) - x_0$ during which
electrons emit at frequency $\nu$ is related to the cooling rate of
the energetic electrons. As this cooling rate is independent on the
electrons initial energy, we can use the result derived in
\S\ref{sec:results1b}, Eq. \ref{eq:t_loss}, to write
$\Delta x = \gamma_j \beta_j c \delta t \approx 6 \pi m_e c^2 \gamma_j
\beta_j / \sigma_T B_0^2 (\gamma \beta)_{\max}$. As the electrons cool
the maximum emission frequency decays. Since the emission frequency
$\nu \propto (\gamma\beta)^2$, emission at frequency $\nu$ can be
obtained only from energetic electrons $(\gamma \beta)_{\max}(x)/
{(\gamma \beta)}_{\min} = (\nu/\nu_{peak,0})^{1/2}$. Using these
results in Eq. \ref{eq:dF_nu_2}, one finds
\beq
F_\nu|_{\nu > \nu_{fast}} = {(m_e c^2)^2
    \over 18 \pi d^2 m_p c \Lambda} {q_j {\dot M}_{\rm disk} \over
      B_0 q \theta_{el,0}} \left( {\nu \over \nu_{peak_0}}
    \right)^{-p/2}.
\label{eq:F_high}
\eeq 

We therefore conclude that for intermediate magnetic field
$B_{cr,0} < B_0 < B_{cr,1}$, the flux above $\nu_{peak,0}$ changes
from $F_\nu \propto \nu^{-(p-1)/2}$ below $\nu_{fast}$, to $F_\nu
\propto \nu^{-p/2}$ at higher frequencies. In reality, we do not
expect a sharp cutoff at $\nu_{fast}$, but a smooth connection (see
Fig. \ref{fig:5}). This can be the source of the steep spectral
slope observed in XTE J118+480 at the X-band, and can affect our
interpretation of the power law index of the accelerated
electrons. See further discussion in \S\ref{sec:summary}.

\subsection{Strong magnetic field}
\label{sec:results2c}

By definition, if $B_0 > B_{cr,1}$ then electrons at the peak of the
Maxwellian, hence all the electrons above the peak rapidly cool by
synchrotron emission close to the jet base.  The results derive in the
previous sections, \S\ref{sec:results1b}, \S\ref{sec:results1c} and
\S\ref{sec:results2b} enable us to obtain the emission in the entire
spectral range without the need for additional calculations.

Since $\nu_{fast} < \nu_{peak,0}$, the flux at $\nu_{peak,0}$ is
given by Eq. \ref{eq:F_nu_peak}, while the flux at higher
frequencies drop as $F_\nu \propto \nu^{-p/2}$. At lower frequencies,
$\nu_{fast} < \nu < \nu_{peak,0}$, the flux is $F_\nu \propto
\nu^{-1/2}$. At even lower frequencies, $\nu < \nu_{fast}$, the flux
depends on the jet geometry, $F_\nu \propto \nu^{1-1/a_{jet}}$. For
intermediate magnetic field, $B_0 < B_{cr,2}$ the results of
\S\ref{sec:results1b} show that at low frequencies $F_\nu \propto
B_0$, while if the magnetic field is strong, $B_0 > B_{cr,2}$ then the
flux at low frequencies is suppressed by self absorption, $F_\nu
\propto B_0^{-8}$ (see \S\ref{sec:results1c}).

\begin{figure}
\plotone{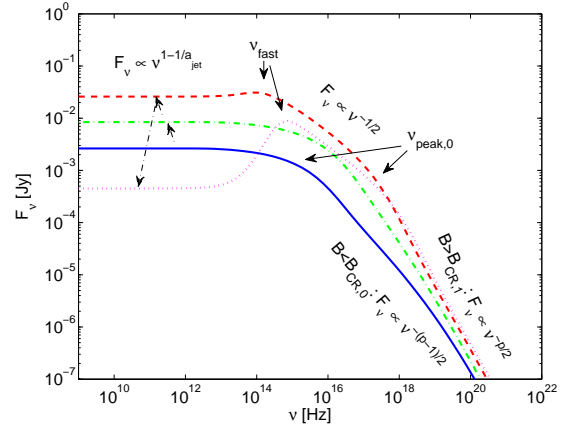}
\caption{Examples of spectra obtained from an initial power law
  distribution of electrons with power
  law index $p=2.5$, for various values of the magnetic field, around
  the critical values $B_{cr,1}$ and $B_{cr,2}$: $B_0 = 3\times
  10^3$~G (solid, blue), $10^4$~G (dash-dotted, green), $10^5$~G
  (dash-dash, red), and $3\times 10^5$~G (Dotted, purple). Adiabatic
  energy losses are neglected here. All the
  other parameters are the same as in Fig. \ref{fig:2}.  The key
  result here is that an increase of the magnetic field above
  $B_{cr,2}$ decreases the radio spectra, with only minor effect on
  the spectra at the X- ray band.  Marked are the transition
  frequencies $\nu_{peak,0}$ and $\nu_{fast}$. For $B_0 > B_{cr,2}$,
  $\nu_{fast}$ increases with $B_0$ (see Eq. \ref{eq:nu_fast2}). The
  dashed arrows indicate the evolution of the spectra with the
  increase of the magnetic field. At low frequencies, below
  $\nu_{peak,0}$, the spectra obtained from Maxwellian distribution
  (\S\ref{sec:results1}) for similar values of the magnetic field is
  similar to that obtained here.  At higher frequencies,
  above $\nu_{peak,0}$ the spectral slope gradually changes from
  $(p-1)/2$ for $B_0 < B_{cr,0}$ to $p/2$ for $B_0 > B_{cr,1}$. }
\label{fig:5}
\end{figure}

Comparison of the flux for different values of the magnetic field is
presented in Fig. \ref{fig:5}. As discussed above, regardless of the
value of the magnetic field, the low energy part of the spectrum,
below $\nu_{peak,0}$ is similar in this scenario and in Maxwellian
distribution scenario discussed in \S\ref{sec:results1}. We thus
conclude that it is not possible to discriminate between these two
initial distributions by observations at low energies, radio - IR
bands. At higher frequencies, above $\nu_{peak,0}$, the spectral slope
gradually changes from $(p-1)/2$ for $B_0 < B_{cr,0}$ to $p/2$ for
$B_0 > B_{cr,1}$. However, as is seen in the figure, this transition
is gradual. We therefore expect an accurate measurement of the power
law index $p$ to be difficult.

\section{Maxwellian distribution of electrons with adiabatic energy
  losses} 
\label{sec:results3}

While in the previous sections we neglected adiabatic energy losses, in
this section we consider these, in addition to the synchrotron
losses. As a result, the electrons temperature do not asymptote to a
constant value, but continuously decreases as the electrons propagate
along the jet (see Eq. \ref{eq:elec_energy1}). This introduces an
additional break, $\nu_{trans}$ (see \S\ref{sec:parms2}, equation
\ref{eq:nu_trans}).

As we will show below, a major part of the analysis carried in the
previous sections holds here as well. This is due to the fact that in
strong magnetic field, the initial rapid cooling of the electrons
close to the jet base, which determines the flux at high energies, is
not affected by the inclusion of adiabatic losses (see discussion in
\S\ref{sec:elec_temporal_behavior}). These only affect the late time
decay of the electrons temperature, hence the emission at low (radio)
frequencies.

\subsection{Weak magnetic field: spectral break at the transition frequency}
\label{sec:results3a}

If $B_0 < B_{cr,1}$, then the initial rapid cooling is insignificant,
and the electrons temperature vary along the jet as $\theta_{el}
\propto x^{-2a_{jet}/3}$ (see Eq. \ref{eq:elec_energy1}). The
magnetic field also decays, and as a result, the peak frequency decays
as $\nu_{peak} \propto x^{-7 a_{jet}/3}$ and the break frequency
decays as $\nu_{thick} \propto x^{-2 a_{jet}/3}$ (see
\S\ref{sec:parms2}, Eq. \ref{eq:var2}). Assuming that at the jet
base $\nu_{peak,0} > \nu_{thick,0}$, the difference in the decay laws
introduces an additional break in the observed spectrum, at frequency
$\nu_{trans}$ (\S\ref{sec:parms2}, Eq. \ref{eq:nu_trans}).

At the frequency range $\nu_{trans} < \nu < \nu_{peak,0}$, one can
estimate the observed flux in a very similar way to the calculations
done in \S\ref{sec:results1a}, by integrating the emission from the
different jet segments. The decay of the peak frequency implies that
contribution to the flux at a given frequency $\nu < \nu_{peak,0}$ is
only from jet regions $x \leq x_{\max}$, where $x_{\max}/x_0 =
(\nu/\nu_{peak,0})^{-3/(7 a_{jet})}$. Integrating the emission from
the jet segments $x_0 \leq x \leq x_{\max}$ while considering only the
optically thin part of the spectrum (see Eq. \ref{eq:F_nu_thin})
gives
 \beq
\ba{lcl}
F_\nu (\nu_{trans} <\nu< \nu_{peak,0}) & \simeq & {2 r_0^2 A_0 \over
  9 d^2} {\theta_{el_0}^3 q^3 B_0 \over m_e c^2} \nonumber \\
& & \times
\int_{x_0}^{x_{\max}} \left({x \over x_0}\right)^{-a_{jet}} \nonumber
\\ & & \times
\left[{3 \nu \over 4 \nu_{peak,0}}\left({x \over
      x_0} \right)^{7 a_{jet}/3}  \right]^{1/3} 
 dx  \nonumber \\
& = & \left({3 \over 4}\right)^{1/3} 
{2 r_0^2 \over  9 d^2}{u_0 q^3 B_0 x_0 \over (1-2a_{jet}/9)
  m_e c^2 m_p c^2} \nonumber \\
& & \times \left({ \nu \over  \nu_{peak,0}}
\right)^{({3/ 7})\left(1-1/a_{jet}\right)}.  
\ea
\label{eq:F_ad_weak}
\eeq
We therefore find that at the frequency range $\nu_{trans} < \nu <
\nu_{peak,0}$, the flux depends on the jet geometry as $F_\nu \propto
\nu^{(3/7)(1-1/a_{jet})}$. This dependence is different than in the
pure synchrotron case considered in \S\ref{sec:results1a}, however we
note that for conical jets ($a_{jet}=1$), still a flat spectrum is
obtained. At frequencies higher than $\nu_{peak,0}$ the flux decays
exponentially for a Maxwellian distribution of electrons.

The main difference between this scenario and the pure synchrotron
scenario discussed in \S\ref{sec:results1a} is the appearance of the
transition frequency, $\nu_{trans}$. This transition frequency is
unavoidable, as it results from the different decay laws of
$\nu_{peak}$ and $\nu_{thick}$ that take place as the electrons
propagate along the jet.  By definition, at frequencies $\nu <
\nu_{trans}$, the peak of the emission is in the optically thick part
of the spectrum ($\nu_{peak} < \nu_{thick}$), and is thus
obscured. The flux at low frequencies $\nu < \nu_{trans}$ is
therefore dominated by emission at $\nu_{thick}$.  This scenario is
similar to the one discussed in \S\ref{sec:results1c}. We
showed there that for $\nu_{peak} < \nu_{thick}$, the evolution of
$\nu_{thick}$ is determined by solving the Eq. $[3
\nu_{thick}(x)/ 2 \nu_{peak}(x)]^{1/3} = (2/3) \log[\pi^2 r(x) A q^2 c
/ 4 \nu_{thick}(x)]$. %An exact solution of this equation can only be
%obtained numerically. However, a
A rough estimate of the flux at low
energies can be done by approximating the logarithm as constant, which
enables to write $\nu_{thick} \propto \nu_{peak} \propto x^{-7
  a_{jet}/3}$.  Since in this case of emission in the optically thick
part of the spectrum, $dF_\nu \propto \nu^2 r \theta_{el} dx \propto
\nu^2 r^{1/3}dx $ (see Eq. \ref{eq:F_nu_thick}), we conclude that
$F_\nu (\nu < \nu_{trans}) \propto \nu^{(13-3/a_{jet})/7}$.
 
\begin{figure}
\plotone{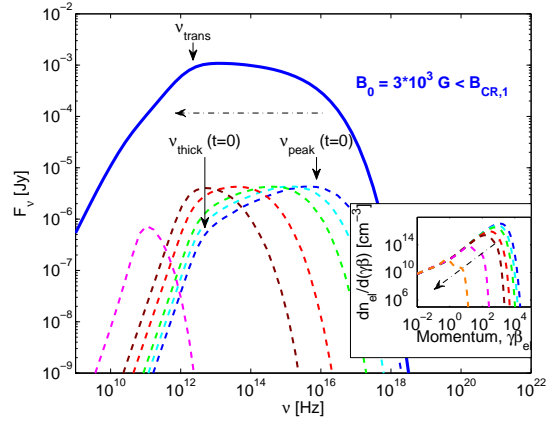}
\caption{Example of the spectra obtained from an initial Maxwellian
  distribution of electrons radiating in weak magnetic field, $B_0
  = 3\times 10^3{\rm \, G} < B_{cr,1}$, when adiabatic energy losses
  are considered. All the other parameters are the same as in figure
  \ref{fig:2}.  The dash-dotted arrows show the direction of the
  evolution of the electrons distribution (inner inset) and flux (main
  panel).  The electrons main cooling mechanism is adiabatic energy
  losses, which results in $\theta_{el} \propto x^{-2 a_{jet}/3}$. The
  transition frequency $\nu_{trans}$ is clearly marked. The
  decomposition into emission from jet segments, illustrates that the
  transition occurs once the emission from the jet segments becomes
  optically thick, above $x_{trans}$ (see equation
  \ref{eq:x_transition}). At $\nu_{trans} < \nu < \nu_{peak,0}$,
  $F_{\nu} \propto \nu^{(3/7)(1-1/a_{jet})}$, which for $a_{jet}=1$ as
  considered here results in flat spectra. However, below
  $\nu_{trans}$ a flat spectra cannot be obtained (see text for
  details).}
\label{fig:6}
\end{figure}

An example of the obtained spectrum is illustrated in figure
\ref{fig:6}. The decomposition shows the evolution of the
characteristic frequencies $\nu_{thick}$ and $\nu_{peak}$ along the
jet. The faster decay of $\nu_{peak}$ is clearly pronounced. We mark
the transition frequency $\nu_{trans}$. The different spectral regimes
are easily identified in this plot.

\subsection{Intermediate and strong magnetic field}
\label{sec:results3b}

For higher values of the magnetic field $B_0 > B_{cr,1}$, the
electrons rapidly cool by synchrotron emission close to the jet
base. Only after the initial rapid cooling, the electrons temperature
decay law asymptotes to $\theta_{el} \propto x^{-2 a_{jet}/3}$. During
the initial rapid decay the adiabatic energy losses are negligible
(see \S\ref{sec:elec_temporal_behavior}, equation
\ref{eq:elec_energy1}). This enables us to use the analysis carried in
\S\ref{sec:results1b} and \S\ref{sec:results1c} in calculating the
flux at high frequencies, which are dominated by emission during the
initial rapid decay phase. On the other hand, the flux at low
frequencies is governed by emission from electrons as they propagate
along the jet. This enables us to use the analysis carried for the
case of weak magnetic field in \S\ref{sec:results3a} in calculating
the flux at low frequencies.

For intermediate magnetic field $B_{cr,1} < B_0 < B_{cr,2}$, similar
analysis to the one carried in \S\ref{sec:results1b} shows the
existence of a break frequency at $\nu_{fast}$, whos value is given in
Eq. \ref{eq:nu_fast}. At the frequency range $\nu_{fast} < \nu <
\nu_{peak,0}$, the flux decays as $F_{\nu} \propto \nu^{-1/2}$, and at
higher frequencies, $\nu > \nu_{peak,0}$, the flux decays
exponentially.

The initial rapid cooling results in a shift of $\nu_{trans}$ to
higher frequency, given by Eq. \ref{eq:nu_trans2}.  As discussed
in \S\ref{sec:frequencies2}, for $B_0 < B_{cr,2}$, $\nu_{trans} <
\nu_{fast}$. Since below $\nu_{fast}$ the analysis carried in
\S\ref{sec:results3a} holds, we conclude that at the range
$\nu_{trans} < \nu < \nu_{fast}$, $F_\nu \propto
\nu^{-(7/3)(1-1/a_{jet})}$, while at lower frequencies $\nu <
\nu_{trans}$, $F_\nu (\nu < \nu_{trans}) \propto
\nu^{(13-3/a_{jet})/7}$.

\begin{figure}
\plotone{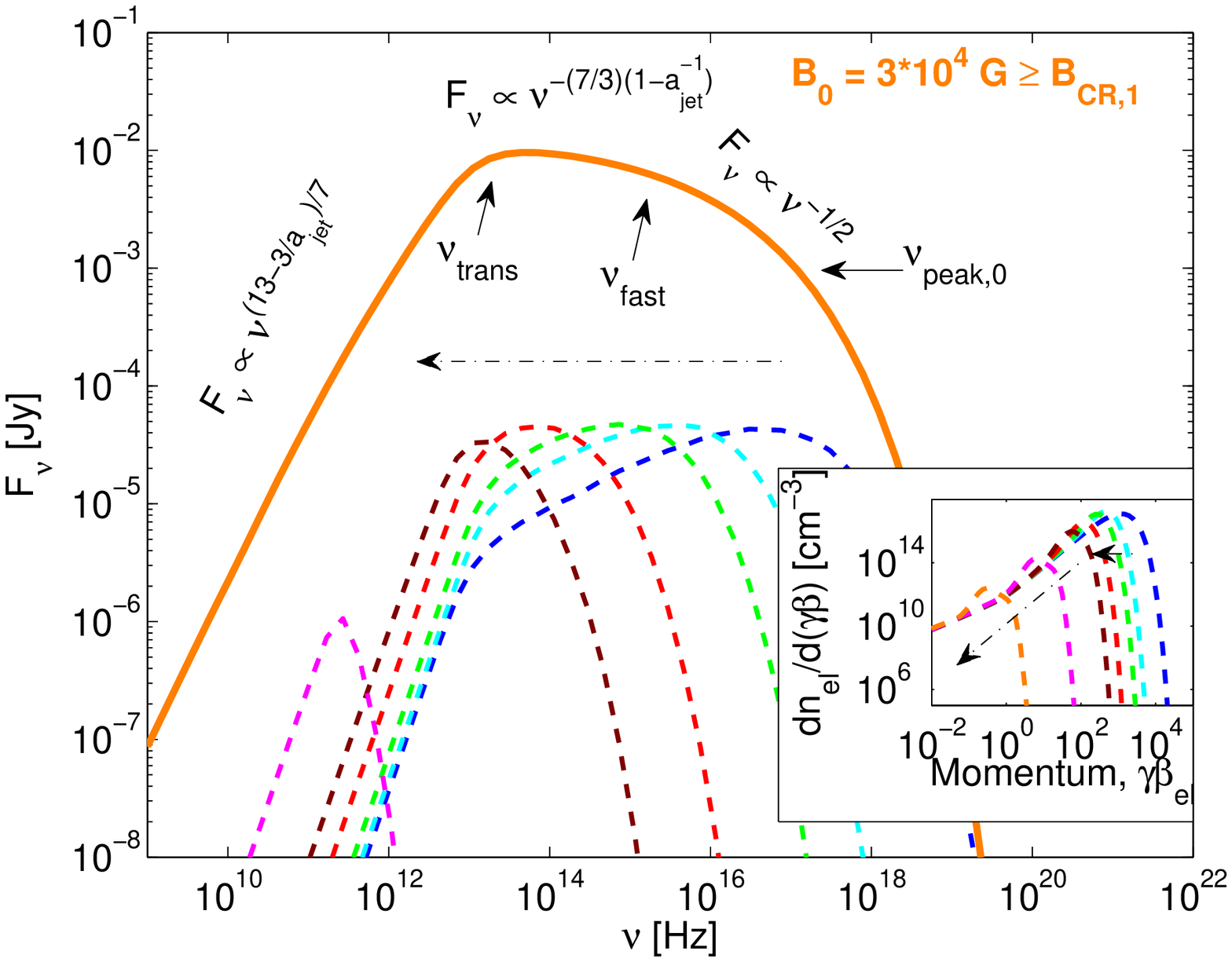}
\caption{Example of the spectra obtained from an initial Maxwellian
  distribution of electrons radiating in intermediate magnetic field,
  $B_{cr,1} \lesssim B_0 = 3\times 10^4{\rm \, G} < B_{cr,2}$, when
  adiabatic energy losses are considered. All the other parameters are
  the same as in Fig. \ref{fig:2}.  The dash-dotted arrows show the
  direction of the evolution of the electrons distribution (inner
  inset) and flux (main panel). The initial rapid decay of the
  electrons temperature is seen in the inner panel.  While we mark in
  this figure the three transition frequencies, $\nu_{peak,0}$,
  $\nu_{fast}$ and $\nu_{trans}$, clearly for intermediate value of
  $B_0$ the transition in the flux behaviour is smooth.  }
\label{fig:7}
\end{figure}

We thus conclude that for $B_{cr,1} < B_0 < B_{cr,2}$, the spectra
below $\nu_{peak,0}$ has three different regimes, separated by the two
transition frequencies $\nu_{trans}$ and $\nu_{fast}$. In practice,
however, we expect these transitions to be smooth, as is illustrated in
Fig. \ref{fig:7}.

\begin{figure}
\plotone{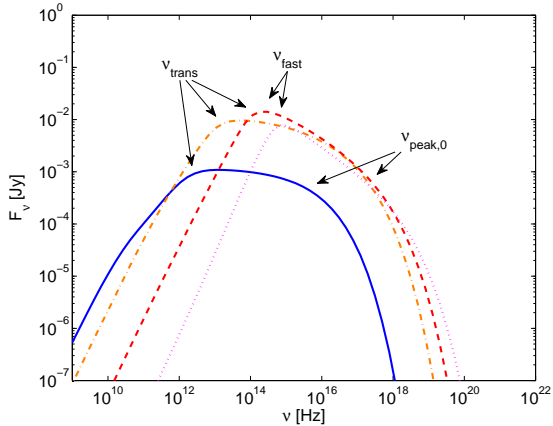}
\caption{Example of spectras obtained from an initial Maxwellian
  distribution of electrons when adiabatic energy losses are included,
  for various values of the magnetic field, around the critical values
  $B_{cr,1}$ and $B_{cr,2}$: $B_0 = 3\times 10^3$~G (solid, blue),
  $3\times 10^4$~G (dash-dotted, orange), $10^5$~G (dash-dash, red),
  and $3\times 10^5$~G (dotted, purple).  All the other parameters are
  the same as in Fig. \ref{fig:2}. As the magnetic field increases,
  $\nu_{trans}$ shifts to higher frequencies, and for $B_0 \geq
  B_{cr,2}$, $\nu_{trans} = \nu_{fast}$. As a result, a flat radio
  spectra cannot be obtained at low frequencies in this model.  The
  increase of the magnetic field results in a shift of $\nu_{fast}$ to
  higher frequencies, as predicted by Eq. \ref{eq:nu_fast2}. For $B_0
  > B_{cr,1}$, the flux at higher frequencies is nearly insensitive to
  the exact value of the magnetic field.  }
\label{fig:8}
\end{figure}

If the magnetic field is stronger, $B_0 > B_{cr,2}$, then, by
definition emission below $\nu_{fast}$ is in the optically thick part
of the spectrum. In this case, $\nu_{trans} = \nu_{fast}$ (see
discussion in \S\ref{sec:frequencies2}).  Thus, the spectrum below
$\nu_{peak,0}$ in this case is composed of only two segments: $F_\nu
\propto \nu^{-1/2}$ for $\nu_{fast} < \nu < \nu_{peak,0}$, and $F_\nu
\propto \nu^{(13-3/a_{jet})/7}$ at lower frequencies. This is
illustrated in Fig. \ref{fig:8}.  We note that in this scenario, the
requirement for a flat radio spectra ($F_\nu \propto \nu^0$ below
$\nu_{trans}$) can not be fulfilled for $a_{jet}>1/2$.

\section{Power law distribution of electrons with adiabatic energy
  losses} 
\label{sec:results4}

When adiabatic energy losses are included, the electrons power law
distribution above the Maxwellian affects not only the high energy
part of the spectrum ($\nu > \nu_{peak,0}$), but also the very low
energy part, $\nu < \nu_{trans}$ as well. This is due to the inclusion
of $\nu_{low}$, as discussed in \S\ref{sec:frequencies2}.

The analysis of the spectra in this case is very similar to the
analysis already done in the former sections. The observed spectra is
composed of several break frequencies, in accordance to the value of
the magnetic field.

For weak magnetic field, $B_0 < B_{cr,0}$, the spectra is composed of
four distinctive segments, separated by three break frequencies:
$\nu_{peak,0}$, $\nu_{trans}$ and $\nu_{low}$, given by Eqs.
\ref{eq:nu_p} , \ref{eq:nu_trans} and \ref{eq:nu_low}. Above
$\nu_{peak,0}$ and below $\nu_{\max,0}$, the flux $F_\nu \propto
\nu^{-(p-1)/2}$ is given by Eq. \ref{eq:F_tot_4} in
\S\ref{sec:results2a}.  At the frequency range $\nu_{trans} < \nu <
\nu_{peak,0}$, the results derived in \S\ref{sec:results3a} hold, and
$F_{\nu} \propto \nu^{(3/7)(1-1/a_{jet})}$.

At lower frequencies, $\nu_{low} < \nu < \nu_{trans}$, the flux can be
calculated in the following way. We showed in \S\ref{sec:frequencies2}
that the spectrum in this range is determined by emission from a power
law distribution of electrons at the break frequency $\nu_{thick} $
which marks the transition from the optically thin to the optically
thick part of the spectrum.  In the optically thick part of the
spectrum (Eq. \ref{eq:F_nu_thick}), $dF_{\nu} \propto r^2
(j_{\nu}/\tau{\nu}) dx \propto r B^{-1/2} \nu^{5/2} dx$, where
$j_{\nu}$ and $\tau_{\nu}$ are the emissivity and optical depth for
power law distribution of the emitting electrons (see
\S\ref{sec:frequencies2}). Using $\nu=\nu_{thick}$, and the dependence
of $\nu_{thick}$ on the position along the jet as derived in
\S\ref{sec:frequencies2}, $\nu_{thick} \propto r^{-A_1}$, one finds
\beq
dF_{\nu} \propto \nu^{5/2 - \left({1 \over A_1}\right) \left( {3
      a_{jet} + 2 \over 2 a_{jet}} \right)},
\label{eq:dF_last} 
\eeq
where $A_1(p) = (8+7p)/(3p+12)$ was derived in
\S\ref{sec:frequencies2}. The result in Eq. \ref{eq:dF_last} is
identical to the result derived by \citet{Kaiser06}; however, we note
here that this result is valid only at the frequency range $\nu_{low} <
\nu < \nu_{trans}$ . Writing $F_{\nu} \propto \nu^{\eta}$ at this
range, we find that for $p=2$ (2.5), for $a_{jet} = 1$, $\eta = 0.45
\, (0.59)$, for $a_{jet} = 3/4$, $\eta = 0.18 \, (0.33)$, and for
$a_{jet} = 2/3$, $\eta = 0.05 \, (0.20)$. One can thus conclude that
for $a_{jet} \lesssim 2/3$, a flat spectrum is expected at this range
$\nu_{low} < \nu < \nu_{trans}$, similar to the conclusion in
\citet{Kaiser06}. 
At much lower frequencies $\nu < \nu_{low}$ the discussion in
\S\ref{sec:results3b} holds, and thus we can approximate
$F_{\nu} \propto \nu^{(13-3/a_{jet})/7}$. At high frequencies, $\nu >
\nu_{max,0}$, an exponential decay in the spectrum is expected.

From the discussion in \S\ref{sec:results2}, we find that the
spectrum obtained for intermediate value of the magnetic field,
$B_{cr,0} < B_0 < B_{cr,1}$ differs than the spectrum obtained for
$B_0 < B_{cr,1}$, by the inclusion of $\nu_{fast}$ (equation
\ref{eq:nu_fast}). Since $\nu_{fast} > \nu_{peak,0}$, this inclusion
only affects the energetic part of the spectrum: for $\nu >
\nu_{fast}$, $F_{\nu} \propto \nu^{-(p/2)}$, while the spectrum at
lower frequencies is similar to the spectrum obtained in the weak
magnetic field case derived above. For magnetic field in this range,
the values of $\nu_{trans}$ and $\nu_{low}$ are determined by
Eqs. \ref{eq:nu_trans2}, \ref{eq:nu_low2}, respectively.

For higher values of the magnetic field, $B_{cr,1} < B_0 < B_{cr,2}$,
$\nu_{fast} < \nu_{peak,0}$  and $\nu_{low} = \nu_{trans}$. The
spectrum is thus separated to four distinctive regimes, which are
different than those for the weak magnetic field derived above.
For $\nu > \nu_{peak,0}$, $F_{\nu} \propto \nu^{-(p/2)}$, and at lower
frequencies $\nu_{fast} < \nu < \nu_{peak,s}$, $F_{\nu} \propto
\nu^{-1/2}$. At even lower frequencies, $\nu_{trans} < \nu <
\nu_{fast}$, $F_{\nu} \propto \nu^{(3/7)(1-1/a_{jet})}$, while at the
low end of the spectra, $\nu < \nu_{trans}$, $F_{\nu} \propto
\nu^{(13-3/a_{jet})/7}$.

For even higher values of the magnetic field, $B_0 > B_{cr,2}$,
$\nu_{trans} = \nu_{fast}$ are given by Eq. \ref{eq:nu_fast2}. For
this value of the magnetic field, there are only three distinctive
spectral regimes below $\nu_{max,0}$: $\nu < \nu_{fast}$, $\nu_{fast} < \nu <
\nu_{peak,0}$ and $\nu > \nu_{peak,0}$. The flux in these regimes is
similar to the flux in the equivalent regimes obtained for weaker
magnetic field. 

\begin{figure}
\plotone{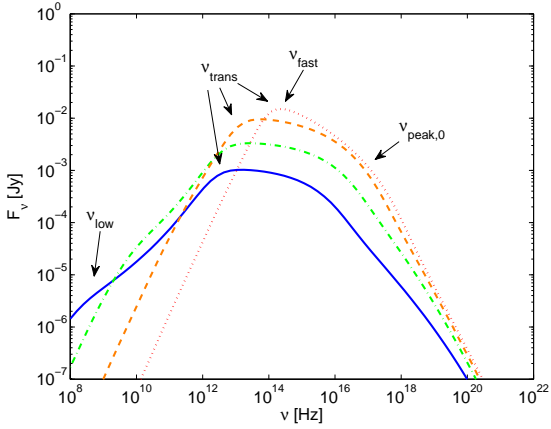}
\caption{Examples of spectra obtained from an initial power law
  distribution of electrons with power law index $p=2.5$ when
  adiabatic energy losses are included, for various values of the
  magnetic field, around the critical values $B_{cr,1}$ and
  $B_{cr,2}$: $B_0 = 3\times 10^3$~G (solid, blue), $10^4$~G
  (dash-dotted, green), $3\times 10^4$~G (dash-dash, orange), and $
  10^5$~G (dotted, red).  All the other parameters are the same as in
  Fig. \ref{fig:2}. We mark the transition frequencies $\nu_{peak,0}$,
  $\nu_{fast}$, $\nu_{trans}$ and $\nu_{low}$. At the frequency range
  $\nu_{low} < \nu < \nu_{trans}$, the flux dependence on the jet
  geometry is described in Eq. \ref{eq:dF_last}. At other frequencies
  the spectral shape is similar to the spectral shape discussed in
  former sections. In particular, the steepening of the slope at high
  frequencies from $(p-1)/2$ to $p/2$ with the increase of $B_0$ is
  pronounced.  }
\label{fig:9}
\end{figure}

\begin{figure}
\plotone{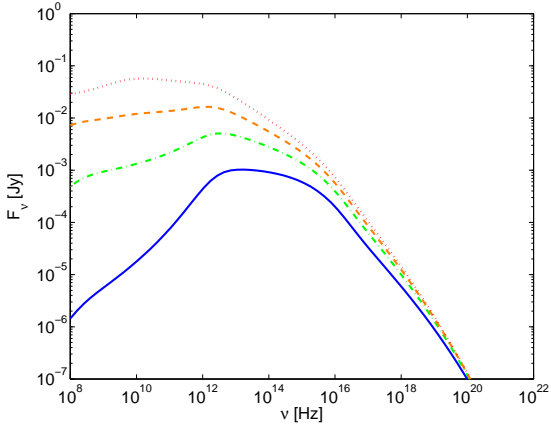}
\caption{Examples of spectra obtained from an initial power law
  distribution of electrons with power law index $p=2.5$ when
  adiabatic energy losses are included, for different jet
  geometries. A conical jet $a_{jet} = 1.0$ (solid, blue), $a_{jet} =
  2/3$ (dash-dotted, green), $a_{jet} = 0.56$ (dash-dash, orange) and
  $a_{jet} = 0.5$ (dotted, red). The magnetic field is taken as $B_0 =
  3\times 10^3 {\rm \; G} < B_{cr,1}$, and all the other parameters
  are the same as in Fig. \ref{fig:2}. For $a_{jet} = 0.56$, the
  analytical result in Eq. \ref{eq:dF_last} predicts a flat spectrum
  $F_{\nu} \propto \nu^0$ at the range $\nu_{low} < \nu <
  \nu_{trans}$. The numerical result shows very good agreement with
  this prediction. }
\label{fig:10}
\end{figure}

Examples of the spectra obtained in this scenario are presented in
Figg. \ref{fig:9} and \ref{fig:10}. In Fig. \ref{fig:9} we present
the dependence of the spectra on the strength of the magnetic
field. We mark the transition frequencies. Note that for strong
magnetic field there is a degeneracy: e.g., as explained above, for
$B_0 > B_{cr,2}$, $\nu_{low} = \nu_{trans} = \nu_{fast}$. In figure
\ref{fig:10} we give examples of the spectra obtained for the several
possible jet geometries. Clearly, for low values of $a_{jet}$, the
flux at low frequencies, at the range $\nu_{low} < \nu < \nu_{trans}$
increases (this holds only as long as $a_{jet} > 1/2$), while the flux
at high frequencies, $\nu > \nu_{peak,0}$ is unaffected. The
theoretical approximation in Eq. \ref{eq:dF_last} predicts that
for power law index $p=2.5$ as is used in the plots, a flat spectra is
obtained for $a_{jet} = 0.56$. The numerical result is in very good
agreement with this prediction.

\section{Narrow jets}
\label{sec:results5}

As discussed in \S\ref{sec:elec_temporal_behavior}, for narrow
jets $a_{jet} < 1/2$, the electrons lose their momentum asymptotically
as $\gamma\beta \propto x^{2 a_{jet} -1}$. This asymptotic decay law
is independent on the inclusion of adiabatic energy losses. Therefore,
the analysis is similar in both scenarios. Moreover, due to the
continuous electrons energy loss along the jet, the analysis is
essentially similar to the analysis carried in \S\ref{sec:results3},
\S\ref{sec:results4} when adiabatic energy losses were included.  We
give here a brief description of the spectrum obtained in this case. As
we will show below, we find it impossible to obtain a flat radio
spectra, as suggested by many observations. We thus find this scenario
less likely.

The high energy part of the spectra, at $\nu > \min(\nu_{peak,0},
\nu_{fast})$ is determined by electrons close to the jet base, and is
therefore unaffected by the jet geometry. The analysis in the previous
sections are thus valid in this case as well.

The difference in the flux emitted in this scenario and the scenarios
discussed above is thus expected at lower frequencies, $\nu <
\min(\nu_{fast}, \nu_{peak,0})$.  The flux at $\nu_{trans} < \nu$ can
be calculated from the decay law of the electrons energy. At this
frequency range, the peak of the synchrotron emission from any jet
segment is in the optically thin part of the spectrum.  As a result,
one can write $\nu_{peak} \propto (\gamma \beta)^2 B \propto x^{3
  a_{jet}-2}$, and using Eq. \ref{eq:F_nu_thin}, $F_{\nu} \propto
r^2 n(r) B dx \propto x^{1-a_{jet}}$. One therefore concludes that at
this frequency range $F_{\nu} \propto \nu^{\eta_1}$, with $\eta_1 =
{(1-a_{jet})/(3 a_{jet}-2)}$. For $a_{jet} =0$, we find $\eta_1 =
-1/2$ while for $ a_{jet} = 1/2$, $\eta_1 = -1$.

If the electrons have a Maxwellian distribution, then at lower
frequencies, $\nu < \nu_{trans}$ the peak of the emitted spectrum is
obscured.  Using Eq. \ref{eq:F_nu_thick}, one finds $dF_{\nu}
\propto \nu^2 r \theta_{el} dx \propto x^{9 a_{jet} -4}$. We have used
here similar arguments to the ones used in \S\ref{sec:results3a}, that
enables us to estimate the $x$-dependence of $\nu_{thick}$ as similar
to that of $\nu_{peak}$ at low frequencies.  We therefore conclude
that at this frequency range, $F_{\nu} \propto \nu^{\eta_2}$, with
$\eta_2 = {(9 a_{jet} -4 )/(3 a_{jet}-2)}$. For $a_{jet} =0$, $\eta_2
= 2$ while for $ a_{jet} = 1/3$, $\eta_2 = 1$. For somewhat wider jet,
$a_{jet} = 4/9$, $\eta_2 = 0$, and a flat spectrum is
obtained. However, as pointed out in \S\ref{sec:parms2}, for
$a_{jet} \approx 0.45$, the transition frequency is very low,
$\nu_{trans} \approx 10^6$~Hz. As a result, the flat part of the
spectra occurs at frequencies well below current observation
capabilities.

If the electrons have a power law distribution above $\gamma_{\min}$,
then the observed spectrum below $\nu_{trans}$ shows an additional
break, at $\nu_{low}$. Similar to the discussion in
\S\ref{sec:results4},  for $\nu < \nu_{low}$, $F_{\nu} \propto
\nu^{\eta_2}$. The flux in the intermediate range, $\nu_{low} < \nu <
\nu_{trans}$ is calculated as follows. An analysis similar to the one carried in
\S\ref{sec:frequencies2} shows that in this case, when writing the
electrons distribution as $n_{el}(\gamma) d\gamma = k(x) \gamma^{-p}
d\gamma$, the proportionality constant evolves as $k(x) \propto
x^{1-4a_{jet} + p(2 a_{jet}-1)}$, resulting in $\nu_{thick} \propto
x^{[a_{jet}(3p-8)+2(1-p)]/(p+4)}$. For $p=2$ this results in
$\nu_{thick} \propto x^{-(1+a_{jet})/3}$, while for $p=2.5$ one
obtains $\nu_{thick} \propto x^{-(6+a_{jet})/13}$. Using $dF_{\nu}
\propto r B^{-1/2} \nu^{5/2} dx$, one finds that for $p=2$, $dF_{\nu}
\propto \nu^{-(1+4 a_{jet})/(2+2 a_{jet})}$, while for $p=2.5$,  $dF_{\nu}
\propto \nu^{(4-34 a_{jet})/(12+2 a_{jet})}$. We thus conclude, that
for $p=2$, the flux varies from $F_{\nu} \propto \nu^{-1/2}$ for
$a_{jet}=0$, to $F_{\nu} \propto \nu^{-1}$ for
$a_{jet}=1/2$. For $p=2.5$ a flat spectrum can actually be obtained at
this range  $\nu_{low} < \nu < \nu_{trans}$, but only for very narrow
jets, $a_{jet} = 0.1$. For such a narrow jet, the analysis carried in
\S\ref{sec:parms2}, \S\ref{sec:frequencies2} shows that $\nu_{low}
\lesssim \nu_{trans}$, and thus the flat spectrum is limited to a very
narrow band, and in practice does not exist.

\begin{figure}
\plotone{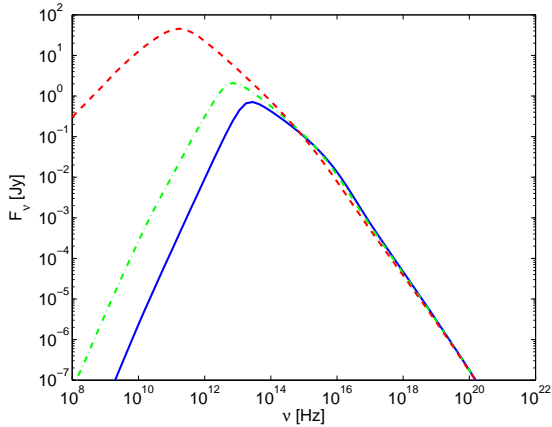}
\caption{Examples of spectra from power law distribution of electrons
  with index $p=2.5$ for different jet geometries in narrow
  jets. Shown are $a_{jet} = 0.1$ (solid, blue), $a_{jet} = 0.2$
  (dash-dotted, green), $a_{jet} = 1/3$ (dash-dash, red). The magnetic
  field is taken as $B_0 = 3\times 10^3 {\rm \; G} < B_{cr,1}$, and
  all the other parameters are the same as in Fig. \ref{fig:2}. The
  transition frequency $\nu_{trans}$ is clearly seen. In no case a
  flat radio spectra can be obtained (see text for details).  }
\label{fig:11}
\end{figure}

Examples of the spectra for the various jet geometries are presented
in Fig. \ref{fig:11}. The transition frequency $\nu_{trans}$ is
clearly seen to evolve to lower frequencies when $a_{jet}$ increases,
as is predicted in \S\ref{sec:parms2}.

\section{Summary and discussion}
\label{sec:summary}

In this paper, we have extensively studied a model for synchrotron
emission from jets in black-hole X-ray binaries. Our basic assumption is that
the electrons are accelerated once at the base of the jet, and lose their
energy by synchrotron emission and possible adiabatic energy losses, as
they propagate along the jet. As the details of the acceleration
process are not understood from first principles, we considered two
scenarios which have strong theoretical and observational motivations:
a Maxwellian distribution of energetic electrons, and a power law
distribution above the low energy Maxwellian, at the energy range
$\gamma_{\min} < \gamma < \gamma_{\max}$, with an
exponential decay at higher energies. The inclusion of a low energy
cutoff to the accelerated electrons energy distribution directly
implies a characteristic break frequency in the observed spectra, at
$\nu = \nu_{peak,0}$ (see Eq. \ref{eq:nu_p}). 
%To the best of our
%knowledge, this break frequency was not discussed so far in the context
%of emission in BHXRBs jets.
This break frequency adds to the
break frequency $\nu_{thick}$ which marks the transition from the
optically thin to the optically thick emission.

We assume that the magnetic field decays along the jet in accordance
to Poynting flux conservation, $B(r) \propto r^{-1}$.  We showed in
\S\ref{sec:elec_temporal_behavior}, that the electrons cooling along
the jet has an analytical description (see Eqs.
\ref{eq:elec_energy1} and \ref{eq:elec_energy2}). These equations hold
the key to the rest of the analysis. By studying the cooling rates, we
found that in strong magnetic field, the cooling can be separated into
two distinctive regimes: an initial rapid cooling that takes place
close to the jet base, during which the electrons emit at high frequencies
(UV, X- and $\gamma$-rays), and a secondary long phase in which the
electrons cooling rate asymptotes. For wide jets, $a_{jet}> 1/2$, when
only synchrotron emission is considered, the electrons energy becomes
time (and $x$) independent, $\gamma \propto x^0$; when adiabatic
energy losses are considered, then $\gamma \propto x^{-2
  a_{jet}/3}$. For narrow jets, both scenarios result in a similar
asymptotic decay law, $\gamma \propto x^{2 a_{jet}-1}$.

These results enabled us to define in \S\ref{sec:B_critical} critical
values of the magnetic field, and in \S\ref{sec:parms2} ,
\S\ref{sec:frequencies2}, five transition frequencies in the observed
spectra. We then analyzed the resulting spectra for the various
assumptions on the accelerated electrons spectra. Maxwellian
distribution was considered in \S\ref{sec:results1} and
\S\ref{sec:results3}, and power law spectra was considered in
\S\ref{sec:results2} and \S\ref{sec:results4}. We further studied the
effect of inclusion of adiabatic energy losses in \S\ref{sec:results3}
and \S\ref{sec:results4}, and the unique scenario of narrow jets,
$a_{jet} < 1/2$, in \S\ref{sec:results5}.

While the variety of spectra that can be obtained in such a simplistic
model is found to be very large, we can point to some general properties
of the spectra, that we find of high importance.

(I) A flat radio spectra, as it is seen in many objects, can be obtained
if only synchrotron emission is considered, for conical jets (see
\S\ref{sec:results1}, \S\ref{sec:results2}, Figg. \ref{fig:2} --
\ref{fig:5}), regardless of the value of the magnetic field. This
scenario is similar to the original model of \citet{BK79},
although there is a different physical origin for the flat spectrum:
here the flat spectrum results from a decay of the peak of the
emission frequency $\nu_{peak}$ along the jet, while in \citet{BK79}
model, the evolution of $\nu_{thick}$ was considered. As we showed in
\S\ref{sec:parms2}, neglecting adiabatic energy losses, both
frequencies evolve in a similar way. We further showed that when
adiabatic energy losses are included, a flat spectrum can be obtained
for conical jets only above $\nu_{trans}$, whose value is given in
Eq. \ref{eq:nu_trans} (see \S\ref{sec:results3}, \S\ref{sec:results4},
Figg. \ref{fig:6} -- \ref{fig:9}). At lower frequencies, a flat
spectrum can be obtained only when a series of conditions are met:
the electrons are power law distributed, the jet has a specific
geometry (see \S\ref{sec:results4}, equation
\ref{eq:dF_last} and Fig. \ref{fig:10}), and the magnetic field is
limited. In this case, a flat spectrum is obtained only at the range 
 $\nu_{low} < \nu < \nu_{trans}$. For narrow jets, we showed
in \S\ref{sec:results5} that a flat radio spectra cannot be achieved
(see Fig. \ref{fig:11}).

(II) We showed that the flux at the radio wavelengths depends on the
value of the magnetic field in a non-trivial way: for magnetic field
at the jet base $B_0 < B_{cr,2}$, the flux increases with the increase
of the magnetic field (see \S\ref{sec:results1a}). However, for
stronger magnetic field, $B_0 > B_{cr,2}$ a further increase of the
magnetic field at the jet base leads to a rapid decrease of the
observed radio flux, due to an increase in the optical depth (see
\S\ref{sec:results1c}). This is not accompanied by a similar change in
the flux at higher frequencies, $\nu > \nu_{fast}$: the flux at high
frequencies asymptotes to a constant value in a strong magnetic field,
$B>B_{cr,1}$ .  We therefore find a natural mechanism that can lead to
a variation in the ratio of the radio to X ray fluxes, by a simple
change in the value of the magnetic field. We further investigate the
consequences of this idea in a forthcoming paper [Casella \& Pe'er, in
preparation].  These results imply that the strongest radio emission
occurs for $B_0 \approx B_{cr,2}$, or about three orders of magnitude
below equipartition value. While the origin of the magnetic field in
jets is not understood from first principles, we believe that this
value could be used as a guideline for models of magnetic field
production near the jet base.

(III) We showed that for intermediate values of the magnetic field $B_0
> B_{cr,0}$, the flux at X-ray wavelength gradually changes from
$(p-1)/2$ at low energies to $p/2$ at higher energies, due to the
rapid cooling of the most energetic electrons (see
\S\ref{sec:results2b}). For an even higher value of the magnetic
field, $B_0 > B_{cr,1}$, we found that at the range $\nu_{fast} < \nu
< \nu_{peak,0}$, which is typically in the optical-to-X-ray band, the
flux decays as $F_{\nu} \propto \nu^{-1/2}$ (see
\S\ref{sec:results1b}). This decay law occurs close to the jet base,
and is therefore independent on the jet geometry.

This gradual transition of the flux implies that every attempt to
determine the power law index $p$ of the accelerated electrons by
fitting X-ray data should be done with great care. The result $F_{\nu}
\propto \nu^{-1/2}$ above $\nu_{fast}$ is very robust, as it is
independent on the initial distribution of the electrons, or on the
exact value of the magnetic field. It can easily be misinterpreted as
due to synchrotron emission from a power law distribution of
electrons, whose cooling is insignificant. Under this interpretation,
one would come to the wrong conclusion that the power law index $p$ of
the accelerated electrons is $p=2$.  Similarly, at higher frequencies,
the gradual change of the spectral slope from $(p-1)/2$ to $p/2$ as
predicted here, if fitted using a single power law over a limited
band, can lead to a wrong conclusion about the value of the power law
index $p$.  This may be the source of the discrepancy between
measurements of the index $p$ in BHXRBs and in other objects, as
discussed in \S\ref{sec:intro}.

%\bf{these last 10 lines are a bit confusing, starting from "The robustness..", but I can't find a better rephrasing}

(IV) We obtained high radio flux, $\sim 10$~mJy for parameters that
characterize emission from XTE J1118+480, in models of wide jets. This
is similar flux to that observed in this object \citep{Hynes00}. While
we did not aim to fit data in this manuscript, we find these results
encouraging. We further point out that the terms ``wide'' and
``narrow'' jets used here can be somewhat misleading, since they refer
only to the confinement of the jet. A ``wide'' jet as defined here can
be geometrically very narrow. 

Our model is of course far from being able to describe the full
physical processes that are expected to occur inside the jets. Several
radiative processes, like Compton scattering, pair production or the
full effect of synchrotron self absorption on the electrons energy
distribution, are not considered here (see
\S\ref{sec:numerics}). However, these phenomena have only minor
effects on the resulting spectra under the conditions assumed here
\citep[see, e.g.,][]{Kaiser06}. Moreover, we did not consider in this
work the possible contribution of internal energy dissipations (e.g.,
internal shocks) that can lead to multiple accelerations episodes of
electrons as they propagate along the jet.  

%{\bf THE BRIGHT SPOTS YOU MENTION (SEE THE FEW LINES COMMENTED BELOW) ARE FROM A DIFFERENT TYPE OF JETS, NOT THE STEADY ONE WE ARE DISCUSSING HERE. IN PRINCIPLE WE COULD MENTION THE VARIABLE ACCRETION FLOW (WHICH SUGGESTS THAT ALSO THE JET IS VARIABLE), BUT I WOULD SUGGEST NOT TO MENTION ANY REASON WHY THE "MULTIPLE ACCELERATIONS EPISODES" SCENARIO COULD BE REALISTIC, AND INSTEAD WE REFER TO SOME LITERATURE (e.g. KAISER; JAMIL; SPADA; GHISELLINI; ETC...)}

%This is a very likely
%scenario given the bright spots that were observed inside jets in
%several objects \citep[e.g., the superluminal motion observed in
%GRS1915+105; see, e.g.,][]{Fender99}. 

Multiple acceleration episodes of electrons along the jet results in a
complicated spectra, whose details depend on the details of the
acceleration processes (e.g., position, fraction of particles that are
being accelerated, strength of the magnetic field at the acceleration
sites, etc.). The spectra obtained in this work can therefore be viewed
as a basic ingredient of the spectra that results from such multiple
accelerations: spectrum in the more complicated case can be obtained
by a composition of the spectra presented here. In this manuscript, we
focused on the spectra that results from a single acceleration
episode, in order to demonstrate the key physical processes that
occur in the plasma. While our numerical model can
very easily be generalized to include multiple acceleration episodes,
due to the expected complexity of the spectra in this case,
we leave this for a future work. We do stress though, that  any model
that considers particle acceleration to high energies should treat
this phenomenon separately than synchrotron and adiabatic cooling,
since these phenomena have different physical origin.

One of the key uncertainties in models of emission from jets, when
internal dissipation of energy takes place, involves the origin of the
magnetic field. Here, we assumed that the magnetic field originates at
the core, and evolves according to Poynting flux conservation
law. However, an alternative scenario may be that the magnetic field
is produced by the internal shock waves.  In such a scenario, the
magnetic field would evolve in a different way, thus some of the
results derived here would not hold. Such a scenario was considered by
\citet{Kaiser06}, where however a low energy cutoff in the accelerated
electrons distribution was not included, which has a strong influence
on the obtained results.  A full treatment of the spectral dependence
on the origin of the magnetic field is left for future work.

We discussed in this work emission from jets in the low/hard state of
BHXRBs candidates, since in this state there are good evidence for the
existence of jets. However, the physics of emission from jets in other
sources, like AGN's is most probably very similar. We thus expect that
many of the results found here should be relevant for jets in other
astronomical sources as well.

\appendix
\section{Numerical code}
\label{sec:numerics}

The numerical code is based on the code constructed by \citet{PW05b}
in the study of emission from GRB's. The particles are assumed to be
accelerated at the base of the jet. A fraction $0 \leq \epsilon_{pl}
\leq 1$ is assumed to have a power law distribution between $\gamma_{\min}$
and $\gamma_{\max}$, while $1-\epsilon_{pl}$ of the particles have a
low energy Maxwellian tail below $\gamma_{\min}$. The particles energy
distribution is discretized in momentum space ($\gamma \beta$) in
order to obtain accurate calculations when the electrons cooling is
significant.  The code gets as an input the parameters of the flow
(see \S\ref{sec:frequencies}), and calculates the number and energy
densities of the electrons at the jet base.

The code divides the jet into segments, and evolves the magnetic field
at each segment, in accordance to $B(x) \propto x^{-a_{jet}}$. Both
$B_0$ and $a_{jet}$ are considered as free parameters.  Given the
strength of the magnetic field, the code calculates the rate of
electrons cooling as they propagate through each segment (Eq.
\ref{eq:elec_decay}). In parallel, the emitted flux from each segment
(Eq. \ref{eq:dF_nu}) is calculated self consistently by solving
the radiative transfer equation. When calculating the emissivity, the
full cyclo-synchrotron spectra from each momentum bin is calculated
\citep{MNY96, PW05b}. Since the magnetic field varies along the jet, the
emissivity is calculated for a reference value of the magnetic field
and is tabulated. As the magnetic field evolves along the jet, the
characteristic frequencies decay; however, using the scaling law for
the emitted frequencies, $\nu = \nu_0 \times (B/B_0)$ and emitted
power, $P(\nu) = P[\nu_0 \times (B/B_0)] (B/B_0)$ the full emissivity
at any given jet segment is readily obtained by interpolating from the
saved values. Given the emissivity at each segment, the self
absorption coefficient is calculated by solving the
integro-differential equation  \citep[][eq. 6.50]{Rybicki79}, which is
correct for a general distribution of electrons. Calculation of the
optical depth followes the assumption that the line of sight is
perpendicular to the jet axis. 

When adiabatic energy losses are considered, Eq. \ref{eq:ad1},
which is accurate for any value of the electrons momentum is used. The
discrete electrons momentum levels are kept constant. Therefore,
calculation of the electrons cooling is done in two steps. The first
step is to obtain their new momentum, by solving equation
\ref{eq:ad1}. At a second step, the new electrons distribution is
fitted to the original, discrete cells. We use a second order fit in
logarithmic space of both the electrons number and their momentum, which is
proven to conserve both number and energy density.

In this version of the code we do not assume any further interactions
between the photons and the electrons (e.g., Compton scattering,
increase of the electrons energy due to synchrotron self absorption,
pair production, etc.). We therefore found that a simple first order
integration scheme is sufficient in calculating the evolution of
the electrons energy distribution.

\acknowledgments 
We would like to thank Dipankar Maitra and Sera Markoff for careful
reading of the manuscript and for many useful comments and
discussions. We would further like to thank Ed van den Heuvel, Rob
Fender, Ralph A.M.J. Wijers, Peter M\'esz\'aros, Bing Zhang, Mario
Livio and Martin J. Rees for useful discussions.  AP is supported by
the Riccardo Giacconi Fellowship award of the Space Telescope Science
Institute. This work was partially supported by the Netherlands
Organization for Scientific Research (NWO).

\end{document}